\documentclass[prl,twocolumn, nofootinbib,nobibnotes,floatfix,tightenlines,superscriptaddress,notitlepage]{revtex4-1}
\usepackage{epsfig, dsfont, amssymb, amsmath, amsthm, amsfonts, amsbsy, mathrsfs, bm, tabularx, graphicx} 
\usepackage{multirow,float,setspace,ragged2e,upgreek, tikz, mathtools, siunitx}
\usepackage{color, enumitem, hyperref, cleveref, caption, gensymb}
\usepackage{lipsum}

\hypersetup{hidelinks}

\DeclareCaptionJustification{justified}{\justifying}

\usepackage[a4paper, total={7.5in, 10in}]{geometry}

\mathchardef\mhyphen="2D
\let\Omega\varOmega

\begin{document}

\title{Predicting and controlling nonlinear neuro-mechanical locomotion dynamics}

\author{Alexander E. Cohen}
\affiliation{Department of Chemical Engineering, Massachusetts Institute of Technology, Cambridge, MA 02139, USA}
\affiliation{Department of Mathematics, Massachusetts Institute of Technology, Cambridge, MA 02139, USA}
\author{Jörn Dunkel}
\email[Corresponding author:]{dunkel@mit.edu}
\affiliation{Department of Mathematics, Massachusetts Institute of Technology, Cambridge, MA 02139, USA}

\begin{abstract}
    Neuromechanics aims to understand the link between an animal's neural activity and its physical behaviors. 
    Recent advances in experimental and machine learning techniques enable simultaneous recordings of neural and locomotion dynamics over long time periods and across multiple behavioral transitions in worms, flies, and other organisms.
    These high-dimensional datasets present the challenge of inferring interpretable low-dimensional dynamical models that quantitatively connect neural activity and behavioral dynamics. 
    However, despite major experimental and theoretical progress, there is currently no end-to-end model for predicting locomotion and other behaviors from neural activity.
    Here, we present a theoretical and computational framework for inferring multiscale neuromechanical models from state-of-the-art experimental data. Our data-efficient approach combines interpretable spectral mode representations with Helmholtz-Nambu decompositions and Bayesian inference to identify a predictive stochastic model that converts neural activity time series into behavioral locomotion patterns. We first apply this framework to recently published recordings of neural activity and locomotion in the roundworm \textit{Caenorhabditis elegans}, showing that it accurately describes experimentally observed dynamics. We further demonstrate how the inferred model can be used to predict neural activation patterns for controlling \textit{C. elegans} locomotion in real time, providing a basis for future optogenetic experiments.
    Due to its generic formulation, the framework introduced here is broadly applicable to neuromechanical recordings for a wide range of animal species.
\end{abstract}

\maketitle

Neural control of motion and other animal behaviors occurs across different length and time scales, from short-time shape changes enabling locomotion to long-time states of stereotyped activities. 
Understanding the link between neural activity, motion, and behavior is often considered one of the most important problems in neuroscience~\cite{haspel2023time, krakauer2017neuroscience, flavell2020behavioral}. 
Inferring dynamical systems that accurately link neural activity to changes in shape remains a major challenge and has been investigated across various model organisms, ranging from worms and jellyfish to mice and nonhuman primates~\cite{atanas2023brain, schneider2023learnable, kennedy2024dynamics, vinograd2024causal, weinreb2024keypoint, international2023brain, gosztolai2025marble}.
This problem is non-trivial both from a theoretical and an experimental perspective: the concept of behavior is hard to quantify; animal locomotion is complex to model reliably for different species and in different environments; and neural activity is difficult to measure and even the best currently available datasets consist of partial and noisy neuron measurements~\cite{cermak2020whole, wiltschko2020revealing, wiltschko2015mapping, buchanan2017quantifying}.
However, recent experimental advances in live-imaging technology provide simultaneous measurements of behavior, motion, and neural activity, which present the challenge of developing low-dimensional descriptions and predictive models that incorporate all three variables~\cite{atanas2023brain, kato2015global, hallinen2021decoding}. 
Such models would not only enable the prediction of shape dynamics from control inputs such as neural activity or environmental stimuli, but also help to infer neural activity from shape dynamics.
\par
Animal behavior has been extensively studied over the last decades~\cite{flavell2020behavioral} by combining various theoretical approaches, including probabilistic models to identify distinct behavioral regimes~\cite{wiltschko2015mapping,cermak2020whole,buchanan2017quantifying} and adopting concepts from dynamical~\cite{costa2019adaptive,ahamed2021capturing,berman2014mapping} and stochastic systems theory~\cite{costa2024markovian,ronceray2024learning,bruckner2020inferring,bruckner2020disentangling} to identify characteristic short and long time scale behaviors. Since behavioral processes involve a large number of biological, chemical, and physical variables, a key first step in formulating dynamical models is to identify interpretable low-dimensional representations for behavioral and neural dynamics. 
Previous pioneering studies~\cite{stephens2008dimensionality, berman2014mapping, wiltschko2015mapping} demonstrated that animal postures can be efficiently represented by applying principal component analysis (PCA) to data collected for specific animals and environmental conditions. 
While PCA-based models achieve optimal compression for a specific experimental setup, their generalization to other animal species and new environments can be limited. 
Another practical challenge arises from the fact that behavioral transitions often depend on \lq hidden\rq{} variables, such as hormone and neurotransmitter levels~\cite{cermak2020whole} that typically cannot be measured in experiments. 
Important recent work has shown how some of these obstacles can be overcome by considering locally linear models~\cite{costa2019adaptive,cohen2023schrodinger}, time delay embeddings~\cite{costa2024markovian,cermak2020whole,ostrow2024delay}, or latent-variable formulations~\cite{linderman2019hierarchical,lee2024switching}, but many of those studies focused separately on animal behavior or neural activity.
Thus, despite such major theoretical advances, it has remained challenging to formulate a generalizable low-dimensional modeling framework that can link physical behavior and neural activity in a predictive manner, and can be constrained by currently available experimental data.
\par

Here, we develop and apply a generic multiscale modeling framework that combines spectral mode representations~\cite{romeo2021learning, cohen2023schrodinger, dalmasso20224d, mitchell2023tubular} for animal shape deformation with structured nonlinear stochastic models of locomotion dynamics and behavior.
Starting from a spectral representation of posture, we infer a behavioral model with a gradient component that constrains the system to the correct behavioral manifold of shapes and a curl component, parameterized by Nambu-Hamiltonians~\cite{nambu1973generalized, axenides2010strange}, that specifies the nonequilibrium flow dynamics along the manifold.
These manifolds, and their corresponding dynamics, change as the animal transitions between different states of locomotion.
The behavioral model is then connected to neural activity through a stochastic model that predicts neural activity given behavior in a manner that is robust to the fact that even the best available neural datasets typically only contain a subset of the full set of relevant neurons. 
Importantly, by inverting the neural model, we can not only predict motion from neural activity but also control motion by proposing neural activity patterns to achieve the desired behavioral dynamics.  
To validate this end-to-end framework, we present a detailed application to recently published data~\cite{atanas2023brain} of simultaneously recorded \textit{C. elegans} postures and neuron activities, showing that our model successfully recapitulates the experimentally observed dynamics. Furthermore, as guidance for future optogenetic control experiments, we predict how the locomotion of \textit{C. elegans} can be steered along geometrically complex target trajectories by prescribing activation patterns for a subset of neurons in real time. 
Finally, we illustrate how this framework can be applied to neuromechanical recordings for other animal species (Supplementary Section XV). 

\section*{Theoretical and computational framework}

\begin{figure*}[t]
\includegraphics[]{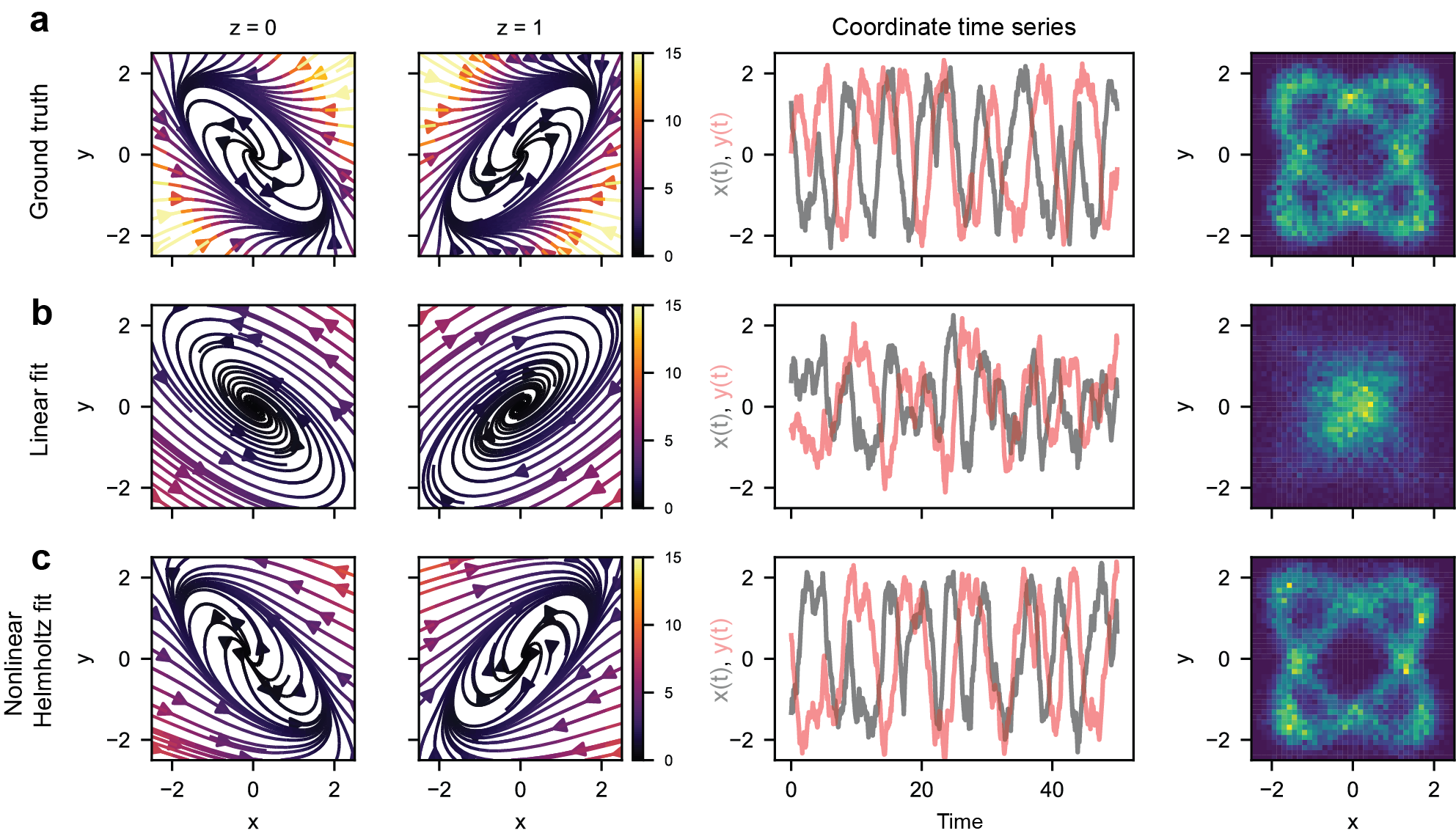}
\caption{\label{fig:fig1} 
Nonlinear potential term corrects the long-term behavior of a two-state system of counter-rotating elliptical limit cycles.
The Helmholtz-Nambu decomposition of this system is given in Supplementary Section XII.
(a) (left) Phase portrait of the deterministic component of the dynamics shows limit cycle behavior in each of the two elliptical states for the ground truth dynamics.
(middle) Ground truth dynamics oscillate in time, reversing phase upon a change in latent state.
(right) Stationary distribution has maxima along the limit cycle.
(b) (left) The linear model captures the general periodic elliptical shape of the dynamics but fails to capture the stable periodic orbit. 
(middle) Thus, there is no constraint on the amplitude of the dynamics and (right) the stationary distribution is approximately a Gaussian mixture.
(c) Explicitly inferring the irrotational and rotational terms separately produces (left) limit cycle-like phase portraits, (middle) amplitude constrained dynamics, and (right) the correct stationary distribution structure.
} 
\end{figure*}

We start by outlining the main theoretical ideas and computational techniques that underlie each modular component of the inference framework. 
Practically speaking, our approach is guided by the typical data limitations of state-of-the-art neuromechanical recordings, which provide comprehensive posture video data, while the accompanying neural activity measurements are sparser, less complete, and exhibit larger uncertainties. 
To obtain a sufficiently flexible and data-efficient description of behavior-dependent shape dynamics of an organism, we use stochastic differential equations~(SDEs) with Markov switching (MS)~\cite{yuan2004convergence, nguyen2019euler} that can be interfaced with a control framework.
To accommodate different limit-cycle behaviors and leverage recent score matching algorithms~\cite{song2019generative}, we use nonlinear Helmholtz decompositions, building on concepts from Nambu-Hamiltonian dynamics~\cite{nambu1973generalized} originally introduced to study quantization of nonlinear systems with multiple conservation laws.
In addition to being computationally advantageous, such Helmholtz-Nambu (HN) decompositions have the theoretical benefit of yielding dynamical models that are intuitively interpretable as motions along the intersections of certain fundamental manifolds. 

\textbf{Behavioral state transitions}.
Different behavioral states of an organism, such as forward motion, backward motion, and turning, can be parameterized by a discrete latent random process $Z_t\in \{z_1,\ldots, z_S\}$, whose stochastic dynamics can be inferred from the data.
These latent dynamics account for unobserved variables that control animal behavior, such as dopamine~\cite{cermak2020whole}.
For the experimental data considered below, we found that the behavioral dynamics $Z_t$ are approximated by a continuous-time Markov chain~\cite{metz1983continuous,10.1214/14-AOAS803} (CTMC) with a transition rate matrix $Q$, where $Q_{ij} > 0$ represents the rate of transition from state $z_i$ to state $z_j$, and $Q_{ii} < 0$ such that each row sums to zero (Supplementary Section III). 
When necessary, it is straightforward to replace the CTMC with more complex stochastic models~\cite{linderman2016recurrent} for the behavioral dynamics $Z_t$.  

\textbf{Behavior-dependent shape-dynamics}.
To describe the geometric shape deformations of an organism in an interpretable low-dimensional space, we consider a generic spectral representation~\cite{cohen2023schrodinger} in terms of time-varying mode amplitudes $\boldsymbol{\hat{\theta}}(t)=[\hat{\theta}_1(t),\ldots, \hat{\theta}_M(t)]$ (Supplementary Section II). 
For instance, below we will use Legendre polynomial decompositions that efficiently parameterize the experimentally observed postures of  \textit{C. elegans}, but depending on the experimental system at hand other mode decompositions are equally possible~\cite{mitchell2023tubular, romeo2021learning}.  
When the organism is in a specific behavioral state $Z_t=z$, we assume that its shape modes evolve according to the Langevin SDE 
\begin{equation}d\boldsymbol{\hat{\theta}} = \mathbf{f}_z(\boldsymbol{\hat{\theta}})\,dt + \boldsymbol{\Sigma}^{1/2}_z\,d\boldsymbol{W}(t)
    \label{eq:SDEMS}
\end{equation}
where the first term on the right-hand side captures the deterministic part of the dynamics, modeled by the nonlinear vector field~$\mathbf{f}_z(\boldsymbol{\hat{\theta}})$. 
The additional \lq noise\rq{} term accounts for fast fluctuating \lq hidden\rq{} variables, effectively modeled by an $M$-dimensional Wiener process $\boldsymbol{W}(t)$ with state-dependent diffusion matrix~$\boldsymbol{\Sigma}_z$. 
Importantly, for each behavioral state $Z_t=z$, there is a different dynamical vector field $\mathbf{f}_z$ and a different covariance matrix $\boldsymbol{\Sigma}_z$ that have to be inferred from the experimental data. 

\par
The combined dynamics of $Z_t$ and
$\boldsymbol{\hat{\theta}}(t)$ forms an SDE-MS
system~\cite{yuan2004convergence, nguyen2019euler}. 
This class of models encompasses the widely used~\cite{wiltschko2015mapping, lee2024switching} autoregressive hidden Markov models, which often enforce a linear form of $\mathbf{f}_z$ to simplify parameter estimation with closed-form expectation-maximization updates~\cite{lee2024switching}. 
Although linear models are easy to fit and can succeed in predicting short-time dynamics, they are unable to describe limit cycles~\cite{strogatz2024nonlinear}, which are common in neuromechanical and other biophysical systems~\cite{swain2002intrinsic,stiefel2016neurons,mattingly2017design,shao2021collective}. 
In contrast, generic nonlinear models can capture a wide range of dynamical behaviors but may be difficult to interpret and constrain with limited data~\cite{quinn2019chebyshev,apgar2010sloppy}. 
To address this problem, we next discuss how geometric Helmholtz-Nambu decompositions yield an interpretable class of nonlinear models that can be efficiently constrained with limited experimental data.

\textbf{Helmholtz decomposition with Nambu-Hamiltonians.}  
We start by decomposing the vector field $\mathbf{f}_z$ as
\begin{equation}
    \mathbf{f}_z=-\frac{1}{2}\boldsymbol{\Sigma}_z{\nabla \Psi}_z+\mathbf{g}_z,
    \label{eq:vector_field_decomposition}
\end{equation}
where this form is obtained by performing a standard Helmholtz decomposition in diffusion-whitened coordinates and transforming back to the original variables (Supplementary Section V).
Next, we restrict to the subclass of fields $\mathbf{f}_z=-\frac{1}{2}\boldsymbol{\Sigma}_z{\nabla \Psi}_z+\mathbf{g}_z$ satisfying $\mathbf{g}_z\cdot \nabla \Psi_z   = 0$ and $\nabla \cdot \mathbf{g}_z=0$, for which the stationary distribution of each state-conditioned SDE reduces to the Boltzmann form $\rho_z \propto e^{-\Psi_z}$~\cite{giorgini2024response} (Supplementary Section IV). 
The main computational benefit of restricting to models of this form is that it allows us to estimate ${\Psi_z}$ directly from the empirically measured stationary distribution via score matching~\cite{song2019generative}, a popular machine learning method widely used in training diffusion models (Supplementary Section VI).
Intuitively, for first-order overdamped dynamics, the gradient component of $\mathbf{f}_z$ represents the equilibrium component of the dynamics and forces the shape variables $\boldsymbol{\hat{\theta}}$ towards the correct behavioral shape space, corresponding to the manifold of the maxima of the stationary distribution.
\par
The complementary curl component, $\mathbf{g}_z$, describes how states flow along shape manifolds.
The divergence-free field $\mathbf{g}_z$ captures the nonequilibrium part of the shape dynamics~\cite{gnesotto2020learning} and can be estimated from the time-ordered trajectories.
To construct a geometrically interpretable family of curl flows, we parameterize $\mathbf{g}_z$ directly in the original shape-mode coordinates by borrowing ideas from Nambu-Hamiltonian dynamics~\cite{nambu1973generalized} (Supplementary Section VII).
To this end, we recall that an $M$-dimensional dynamical system with $M-1$ Nambu-Hamiltonians, $H_i(x_1,\ldots, x_M)$, is given by~\cite{nambu1973generalized} 
\begin{equation}
    \frac{dx_i}{dt} = \sum_{j,k,\ldots,l} \epsilon_{ijk\ldots l} \frac{\partial H_1}{\partial x_j} \frac{\partial H_2}{\partial x_k} \cdots \frac{\partial H_{M-1}}{\partial x_l}.
    \label{eq:hamiltonian}
\end{equation}
The dynamics generated by the divergence-free field on the right-hand side of Eq.~\eqref{eq:hamiltonian} simultaneously conserves each of the Hamiltonians~$H_i$. 
The level sets of these functions define surfaces in the $M$-dimensional space and the solutions of Eq.~\eqref{eq:hamiltonian} live on the intersections of these surfaces. 
If the $H_i$ are polynomials then the solutions correspond to algebraic varieties.
In the applications to neuromechanical data discussed below, parameterizing $\mathbf{g}_z$ with $1$ quadratic and $M-2$ linear Hamiltonians yielded satisfactory results (Supplementary Section X).
\par
\textbf{Model inference and validation.} 
To learn $\Psi_z$ and $\mathbf{g}_z$ from experimental data, we implemented an inference scheme that fits the gradient and curl components sequentially (Supplementary Section X). 
The nonlinear gradient component is inferred using denoising score matching~\cite{song2019generative}, with data augmentation such that $\mathbf{g}_z\cdot \nabla \Psi_z = 0$ is approximately satisfied.
The curl component and noise covariance $\mathbf{\Sigma}_z$ are learned via a maximum-likelihood estimate of the Euler-Maruyama discretization of the time series data~\cite{picchini2007sde, li2020scalable}. We refer to this scheme as stochastic Helmholtz-Nambu (HN) dynamics. 

\par
Before focusing on the application to neuromechanical data, we briefly illustrate the practical benefits of the above framework by considering a basic two-state model, $Z_t\in \{0,1\}$, with a two-dimensional shape space, \mbox{$\boldsymbol{\hat{\theta}} = [x, y]^\top$}, where the system approaches counterclockwise and clockwise rotating elliptical limit cycles for $Z_t=0$ and $Z_t=1$, respectively~(Fig.~\ref{fig:fig1}a; see Supplementary Section XII for the specific equations). 
In this case, a standard parameter estimation method for autoregressive hidden Markov models~\cite{murphy2022probabilistic} yields a qualitatively incorrect best-fit model~(Fig.~\ref{fig:fig1}b). 
By contrast, the stochastic HN dynamics approach identifies a qualitatively and quantitatively predictive model (Fig.~\ref{fig:fig1}c). 

\textbf{Behavior-dependent neuron model}. 
Neuron dynamics are modeled as a stochastic process using a continuous-time hidden Markov model (HMM). 
In this framework, the system is assumed to switch between a discrete set of latent states according to a Markov process defined by a transition-rate matrix, while the experimentally observed neural activities are noisy emissions whose statistics depend only on the current latent state. 
To remain consistent with the shape model, the HMM describing neural activity is constrained to use the same behavioral state-transition matrix inferred from the shape dynamics. For the emission model, the continuous neural activity and its time derivative, $\mathbf{n}(t) = [n_1(t), \dot{n}_1(t), \ldots, n_N(t), \dot{n}_N(t)]$, are discretized into equally spaced bins spanning $\pm 2.5$ standard deviations.
For each hidden state, the emission probabilities are defined as the normalized histogram of these binned observations, such that the empirical frequencies directly parameterize a categorical distribution (Supplementary Sections VII and XI).

\section*{Learning a neuromechanical model for roundworms}

\begin{figure*}[t]
\includegraphics[]{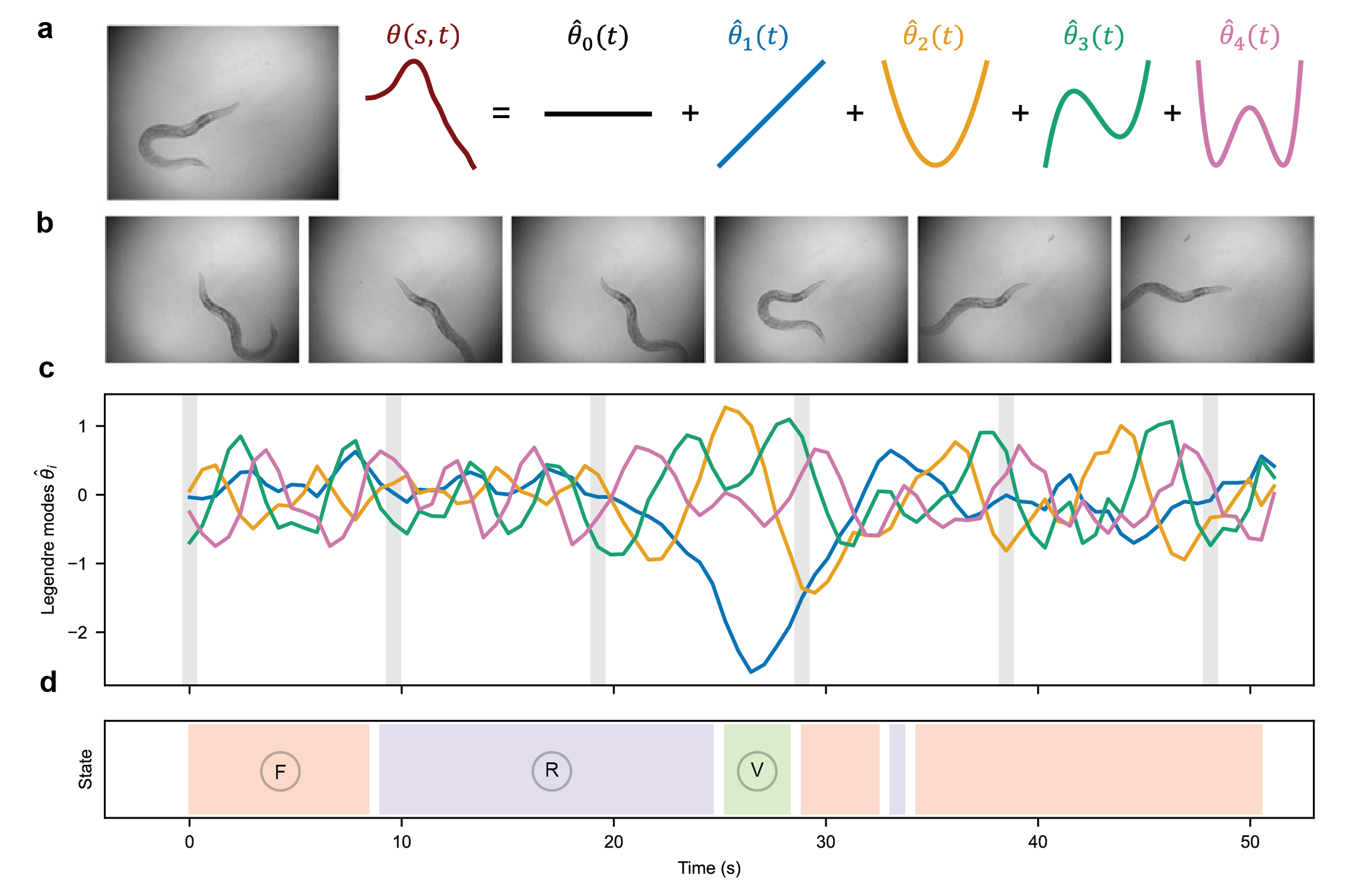}\textit{}
\caption{\label{fig:fig2}
A Legendre polynomial  expansion yields an interpretable,  robust classification of locomotion behaviors via  clustering of mode-amplitude characteristics.
(a)~Worm shape is represented by the centerline's tangent angle $\theta(s,t)$, which changes along the path length $s$ from head to tail  and in time~$t$. Legendre mode coefficients $\hat{\theta}_i(t)$ provide a low-dimensional spectral representation of the posture.
(b) NIR image snapshots during a 50-second worm trajectory, corresponding to the time points indicated by the gray bands below.
(c) Time trace of Legendre modes during the trajectory from the above images. The periodicity of forward and reverse locomotion is captured by the oscillating coefficients with opposite phase for forward versus reverse motion. In addition, spikes of the first-order polynomial coefficient, representing circular curvature, indicate deep body bends.
(d) Behavioral states assigned to the corresponding trajectory segments [forward motion (F), reverse motion (R), ventral turning (V)].} 
\end{figure*}

We now apply the above framework to recent neuromechanical recordings for the roundworm \textit{C. elegans}.

\textbf{Experimental data.}
A recent groundbreaking experimental study~\cite{atanas2023brain} reports simultaneous measurements of posture dynamics and neural activity in freely moving \textit{C. elegans}. 
The study used a spinning disk confocal microscope to measure GCaMP activity, while another light path recorded near-infrared (NIR) bright-field images of the body of a freely moving worm. 
Subsequent NeuroPAL imaging in immobilized worms enables the backmapping of neuron identities to the GCaMP recording, resulting in a dataset of simultaneous \textit{C. elegans} motion and labeled neural activity, of which we use the centerline extracted from the NIR images and the z-scored neural activity traces in our analysis (Supplementary Section I provides more information on data preprocessing and Supplementary Video 1 shows an animation of the raw data used throughout the analysis).
Here, we focus on the four main behavioral locomotion states: forward motion (F), reverse motion (R), dorsal turning motion (D), and ventral turning motion (V) (Fig.~\ref{fig:fig2} and~\ref{fig:fig3}a). 
Similarly, we restrict the activity signal to 12 neurons known to be involved in the control of these behaviors~\cite{atanas2023brain} (Supplementary Section XI and Fig.~\ref{fig:fig4}). 
However, due to the general formulation of the approach, we anticipate that it can be applied to a wider range of behaviors and neurons.

\textbf{Spectral shape representation.}
Recent studies~\cite{romeo2021learning,cohen2023schrodinger,stephens2008dimensionality} have shown that compressed representations of animal posture can be achieved with mode representations.
A widely used representation is based on principal component analysis (PCA). This technique, known as eigenworms~\cite{stephens2008dimensionality} when applied to \textit{C. elegans}, has the strength of optimal compression under a linear dimensionality reduction and has been useful for studying and modeling animal behavior. Here, 
we opt to represent animal shape with a spectral mode representation, which provides three main benefits~\cite{cohen2023schrodinger}:
(i)~the same basis can be used across different animals, mutants, species, and experimental conditions, facilitating comparisons; 
(ii)~the representation is continuous and is computed with exact mathematical transformations, providing easily interpretable physical features of the modes;
(iii)~the basis functions can be chosen to be global, weighted towards specific regions of the animal shape, local, or a combination of these. 

Given that worm posture can be accurately described by a centerline curve, we use orthogonal polynomials, specifically Legendre polynomials.
With the centerline extracted from the microscope images, we obtain curves of the tangent angle along the body coordinate $s$ of the worm over time $t$. These experimentally measured tangent angle fields are expressed in a basis of Legendre polynomials $P_i(s)$ with time-varying coefficients (Fig.~\ref{fig:fig2}a),
\begin{equation}
    \theta(s,t) = \sum_{i=0}^M \hat{\theta}_i(t) P_i(s).
\end{equation}
Legendre polynomials are orthogonal with respect to the weight function $w(s) = 1$, corresponding to a uniform weighting of all body parts.
The 0th order coefficient, associated with $P_0(s) = 1$, represents the average angle along the body, which corresponds to the direction the animal is facing. 
The 1st order coefficient, associated with $P_1(s) = s$, reflects the curvature of the worm's body.
This coefficient tends to spike in magnitude during highly curved postures, such as omega turns, as illustrated in the experimental images and corresponding mode coefficient traces (Fig.~\ref{fig:fig2}b-c). 
Higher-order coefficients capture more wave-like postures and tend to oscillate as the worm undergoes forward and reverse locomotion behaviors, with a change in frequency and phase based on the speed and direction of motion.

\textbf{Behavioral state annotation.}
To identify behavioral states corresponding to forward motions~(F), reverse motions~(R), ventral turns~(V), and dorsal turns~(D), we cluster the coefficient time series, $\hat{\theta}_i(t)$ (Fig.~\ref{fig:fig2}d).
Specifically, we locally fit state-dependent  effective  Hamiltonians~$H_k(\boldsymbol{\hat{\theta}};z)$, $z\in \{F,R,V,D\}$, to the shape mode time series and cluster the time points into distinct states based on the shapes and orientations of the Hamiltonians (Supplementary Video 2, Supplementary Section IX). 
In effect, this method identifies states based on the local geometric structure of the dynamics as encoded by the Hamiltonians.
Other clustering methods for the coefficient time series $\hat{\theta}_i(t)$ yield similar clusters (Supplementary Section IX).

\section*{Results}
We aim to learn a stochastic model for behavior, locomotion, and neural activity in \textit{C. elegans} that generates the conditional distribution of a worm's current behavioral neuro-mechanical state given its past states,  $p(Z(t),\hat{\boldsymbol{\theta}}(t),\mathbf{n}(t) \mid Z(t'),\hat{\boldsymbol{\theta}}(t'),\mathbf{n}(t'))$ for $t'<t$ where,  $Z(t)\in \{F,R,V,D\}$ encodes behavior, $\hat{\boldsymbol{\theta}}(t)$ the shape modes, $\mathbf{n}(t)$ the neural activity time-series. Such a model can provide insights into several important questions in neuroscience by computing various marginal and conditional distributions.
For instance, computing shape given behavior  offers insights into typical postures associated with each behavioral state.
Using the model to compute shape given neural activity elucidates how neural activity translates into motion.
Last but not least, as we demonstrate further below, knowing the distribution of neural activity given shape enables predictions of the patterns of neural activity that  drives the animal along a desired trajectory.

Limited data availability for neural recordings suggests factorizing via Bayes Theorem to separate motion-behavior variables and neural variables (Supplementary Section VIII). 
This factorization allows us to take advantage of the larger number of datasets containing freely moving worms without neural activity and the fact that these datasets typically have a sampling rate more than ten times higher than those with neural activity~\cite{yemini2013database,atanas2023brain}.

\textbf{Learned behavioral model}.
To generate the behavioral distribution, we fit an SDE model with Markov switching of the form in  Eq.~\eqref{eq:SDEMS}, using a Helmholtz-Nambu (HN) split for the dynamical field $\mathbf{f}_z=-\frac{1}{2}\boldsymbol{\Sigma}_z{\nabla \Psi}_z+\mathbf{g}_z$.
The $\boldsymbol{\hat{\theta}}$ dynamics capture short-timescale motion within a state, while the $z$ dynamics capture the longer-timescale behavioral state transitions and effects of unobserved variables such as neurotransmitters.  As anticipated above, we found that a posture representation dimension $M=4$, $S=4$ behavioral states $z\in \{F,R,V,D\}$, and $N=12$ neurons are sufficient to  reproduce experimentally observed statistics and dynamics (Supplementary Section II). 

To learn the behavioral model using the stochastic HN dynamics framework, we geometrically constrain the curl dynamics to lie along a manifold formed by the intersection of equi-surfaces of Hamiltonians $H_k(\boldsymbol{\hat{\theta}};z)$ describing an ellipsoid and two planes, as the mode dynamics typically oscillate with a single dominant frequency (Supplementary Section IX-A).
This constraint yields linear curl dynamics, simplifying the model.
The nonlinear gradient dynamics are learned through denoising score matching and constrained to have maxima along the Hamiltonian intersections, approximately satisfying $\mathbf{g}\cdot\nabla \Psi  = 0$ (see  Supplementary Section X for a detailed description of constraint implementation and  parameter fitting  procedures).

\begin{figure*}[t]
\includegraphics[width=0.8\textwidth]{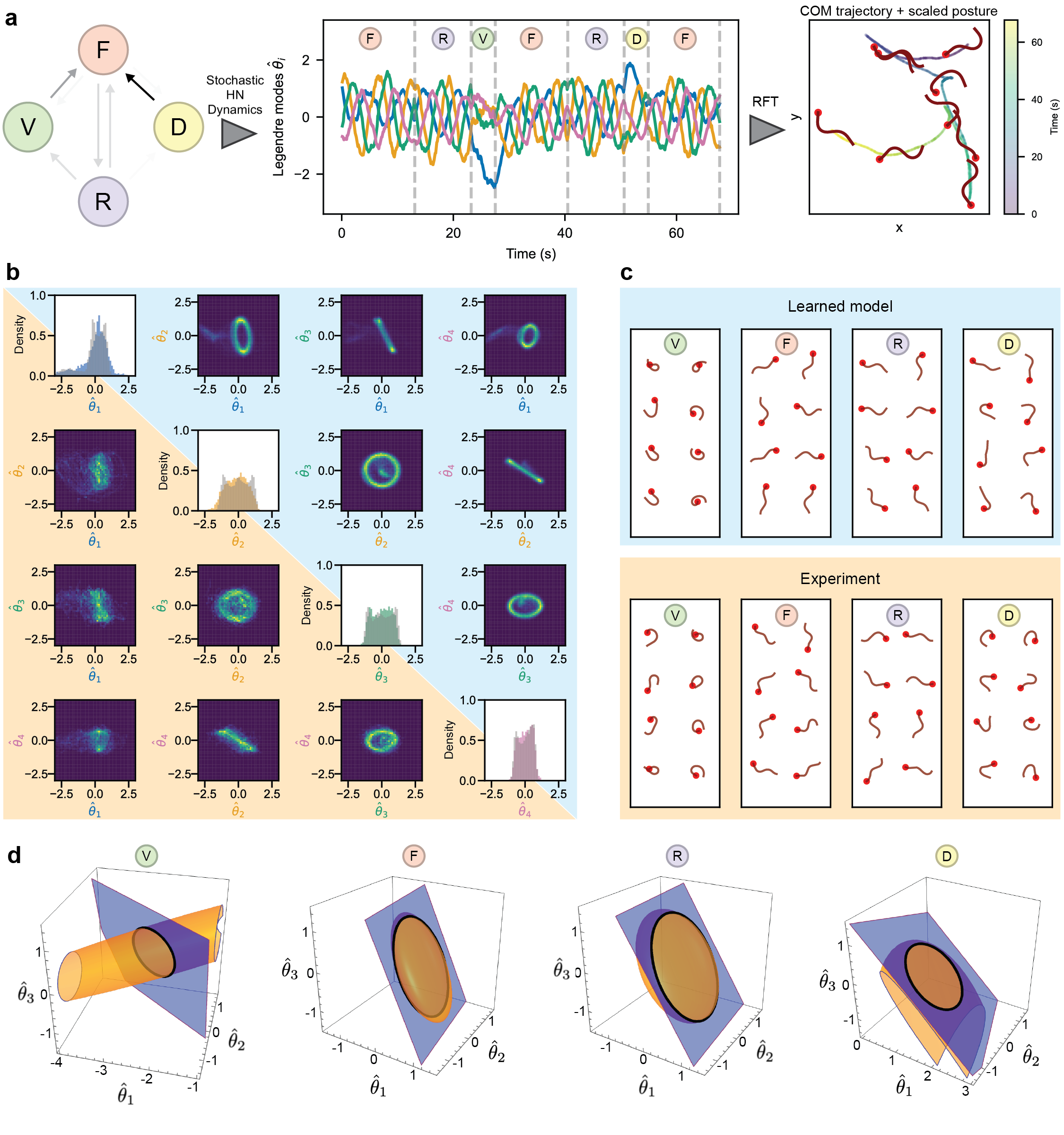}
\caption{\label{fig:fig3} 
The learned model reproduces stereotypical observed worm-shape dynamics and, via a Helmholtz decomposition, separates the gradient (equilibrium) component that generates realistic state-specific postures from the curl (nonequilibrium) component that visualizes the geometric structure of shape dynamics.
(a) Summary of inferred model. (Left) Markov transition rate matrix depicted as a directed, weighted graph. The darkness of the arrows is proportional to the transition rate between the node behavioral states. This Markov transition rate matrix produces a behavioral trajectory, which is combined with the nonlinear model to give a posture trajectory (middle). Here, the behavioral trajectory is chosen to be a stereotypical sequence of a forward-reverse-turn-forward cycle for demonstration. These posture trajectories can be converted to rotational and translational motion with linear resistive force theory.
(b) Stationary distribution from the model (upper right, light blue background) and the experimental data (bottom left, light orange background), represented as pair plots, show similar structure. 
(c) Posture samples from the model's stationary distribution in each behavioral state, learned directly as the irrotational portion of the dynamics. The samples are visually indistinguishable from postures randomly sampled from the experimental data, shown below. The ventral turn state has postures with deep bends along the ventral side and the dorsal turn state has postures with deep bends along the dorsal side. The forward and reverse states have postures with sinusoidal oscillations along the body.
(d) Geometry of the inferred dynamics from the rotational part of the learned model. The surface intersection is centered along large negative values of $\hat{\theta}_1$ in ventral turns and large positive values of $\hat{\theta}_1$ in dorsal turns. The intersections are narrowly centered along zero in $\hat{\theta}_1$ for forward and reverse motion, with circular cross sections along the $\hat{\theta}_2 - \hat{\theta}_3$ plane. The orientations of the surfaces are flipped between forward and reverse motion, giving opposite oscillations.
} 
\end{figure*}

The Markov transition rate matrix governing behavioral switching displays the known dynamics between the four motion states, including the Forward-Reverse-Turn-Forward reorientation cycles and the strong ventral turning bias (Fig.~\ref{fig:fig3}a).
A representative trajectory from this Markov chain drives the stochastic HN dynamics model, which exhibits characteristic oscillatory behavior in the higher, wave-like modes during forward and reverse motion, with spikes in the circular bending modes during the pirouettes (Fig.~\ref{fig:fig3}a, Supplementary Video 3).
After computing the translational and rotational body movements from the posture mode dynamics using resistive force theory~\cite{costa2024markovian,keaveny2017predicting}, the forward and reverse states display a shallow dorsal turn bias (Fig.~\ref{fig:fig3}a), which has also been quantified in freely moving worms~\cite{rozemuller2023statistics}.

Furthermore, the stochastic HN dynamics model accurately reproduces the stationary distribution of the worm dynamics, which distinctly reveals non-Gaussian structures (Fig.~\ref{fig:fig3}b).
Such distributions cannot be captured with purely linear models, which would produce only weighted Gaussian stationary distributions~\cite{hu2020explicit}.
These distributions directly enable generation of worm postures in specific behavioral states, effectively providing the conditional distribution of shape given behavior.
The gradient term in the dynamics corresponds to the score function of the stationary distribution in each state, allowing us to generate representative postures that closely resemble randomly sampled postures from experimental data within each behavioral state (Fig.~\ref{fig:fig3}c).

The geometry of the Legendre-mode dynamics near the attractor is encoded in the intersecting surfaces of the generalized Hamiltonians $H_k(\boldsymbol{\hat{\theta}};z)$ that determine  the curl component of the vector fields (Fig.~\ref{fig:fig3}d, Supplementary Video 4).
Forward and reverse states both exhibit a circular structure centered near zero for each mode with a small radius in the $\hat{\theta}_1$ dimension and radius approximately 1 in the $\hat{\theta}_2$ and $\hat{\theta}_3$ dimensions. 
In contrast, turning states are centered near $\pm 2$ in the $\hat{\theta}_1$ and have smaller radii in the $\hat{\theta}_2$ and $\hat{\theta}_3$ dimensions.
While we have imposed simple geometrical constraints due to limited experimental data, these constraints can be relaxed for more complex systems with larger datasets, and the same analytical framework can be applied.

\begin{figure*}[t]
\includegraphics[]{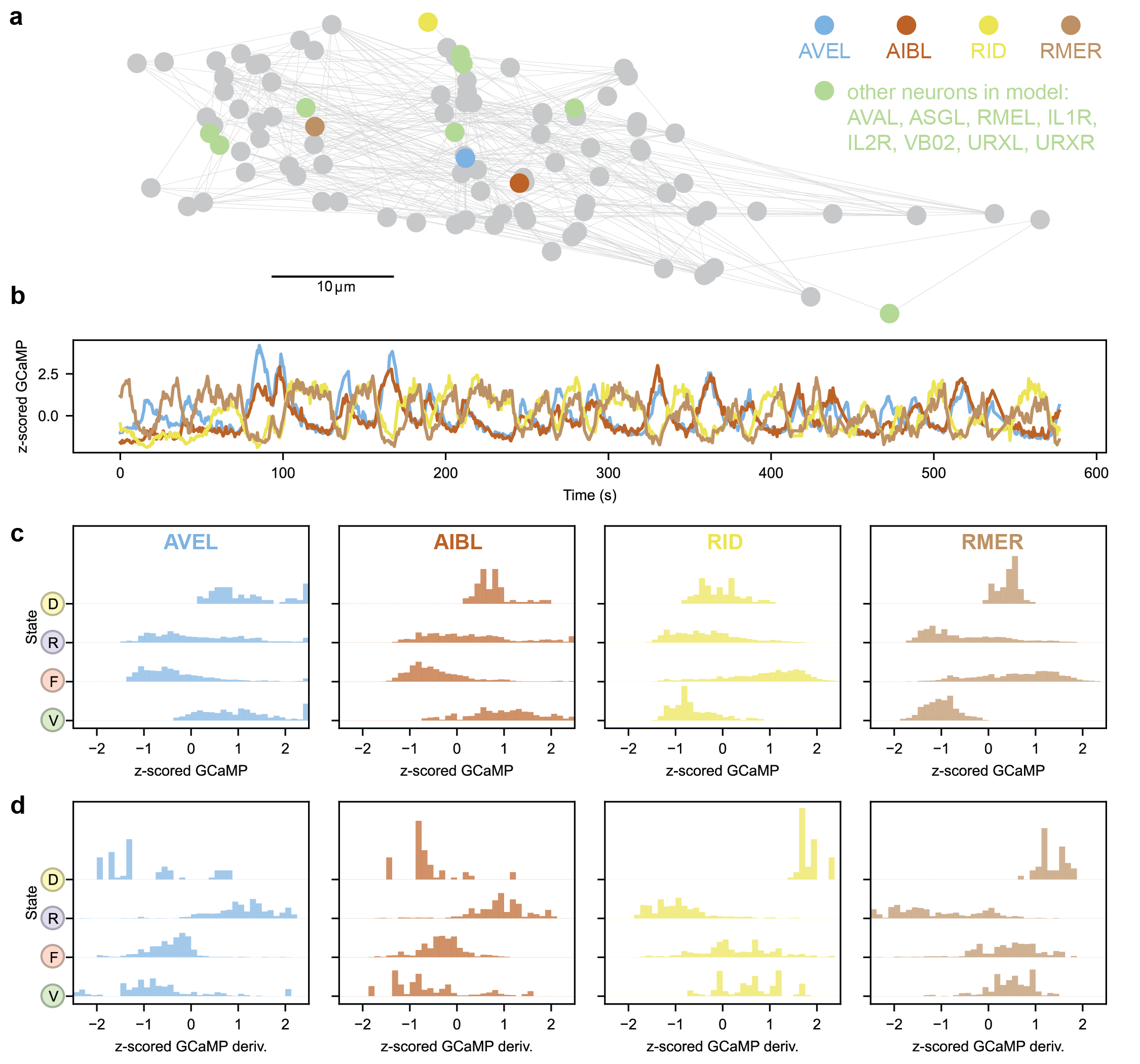}
\caption{\label{fig:fig4} 
Neural activity data used to construct the joint behavior-neural model, including the connectivity of the recorded network, example time traces of neural activity, and the binned distributions of neural activity and its time derivative used for the emission model.
(a) Synaptic connectivity between neurons in the \emph{C. elegans} nervous system, with the neurons positioned anatomically. The neurons used in this analysis are colored. 
(b) A representative trace of neural activity for an example set of 4 neurons during a 10-minute recording. 
(c) Distributions of activity $n_k(t)$ for each of the four example neurons in each behavioral state. Neuron activities tend to display distinct distributions within each behavioral state, which enables the prediction of behavior from neural activity.
(d) Model accuracy can be improved by additionally including statistical information about activity changes, $\dot{n}_k(t)$, as evident from the differences in the marginal distributions. Results and  predictions presented below are based on a model inferred from the  distribution of $\mathbf{n} = [(n_1, \dot{n}_1), \ldots, (n_N, \dot{n}_N)]$ with $N=12$.
} 
\end{figure*}

\textbf{Learned neural model}.
To complete the full model of the worm, we require a model that specifies the neural distribution.
While previous work~\cite{langdon2023unifying,peach2023implicit} on more complex organisms has focused on learning the dynamics of neural circuits and manifolds, the \emph{C. elegans} nervous system is much smaller and the identities of all individual neurons are known.
The connectome of \emph{C. elegans} is also well known~\cite{cook2020connectome,witvliet2021connectomes} (Fig.~\ref{fig:fig4}a), offering the opportunity to learn a full dynamical model based on the physical or functional connectivity~\cite{zhao2024integrative,randi2023neural,shi2023spatial}.
However, since each experiment records only a subset of neuron activity, directly learning a neural model encompassing all neurons is challenging, so we must simplify the problem.
Fortunately, it was shown  recently~\cite{atanas2023brain} that neural activity can be well-approximated by behavioral variables alone, suggesting that individual neurons can be modeled as conditionally independent given behavior. 

The distributions of neural activity and neural derivatives used to build the hidden Markov model are constructed from the time series traces of neural activity (Fig.~\ref{fig:fig4}b, Supplementary Section XI).
These distributions are distinct across different behavioral states for various neurons (Fig.~\ref{fig:fig4}c-d). While in this work we focus on neurons known to be involved in forward, reverse and turning motions, it is straightforward to extend the framework model to include  additional neurons as well as other posture variables like  head curvature. 

\begin{figure*}[t]
\includegraphics[]{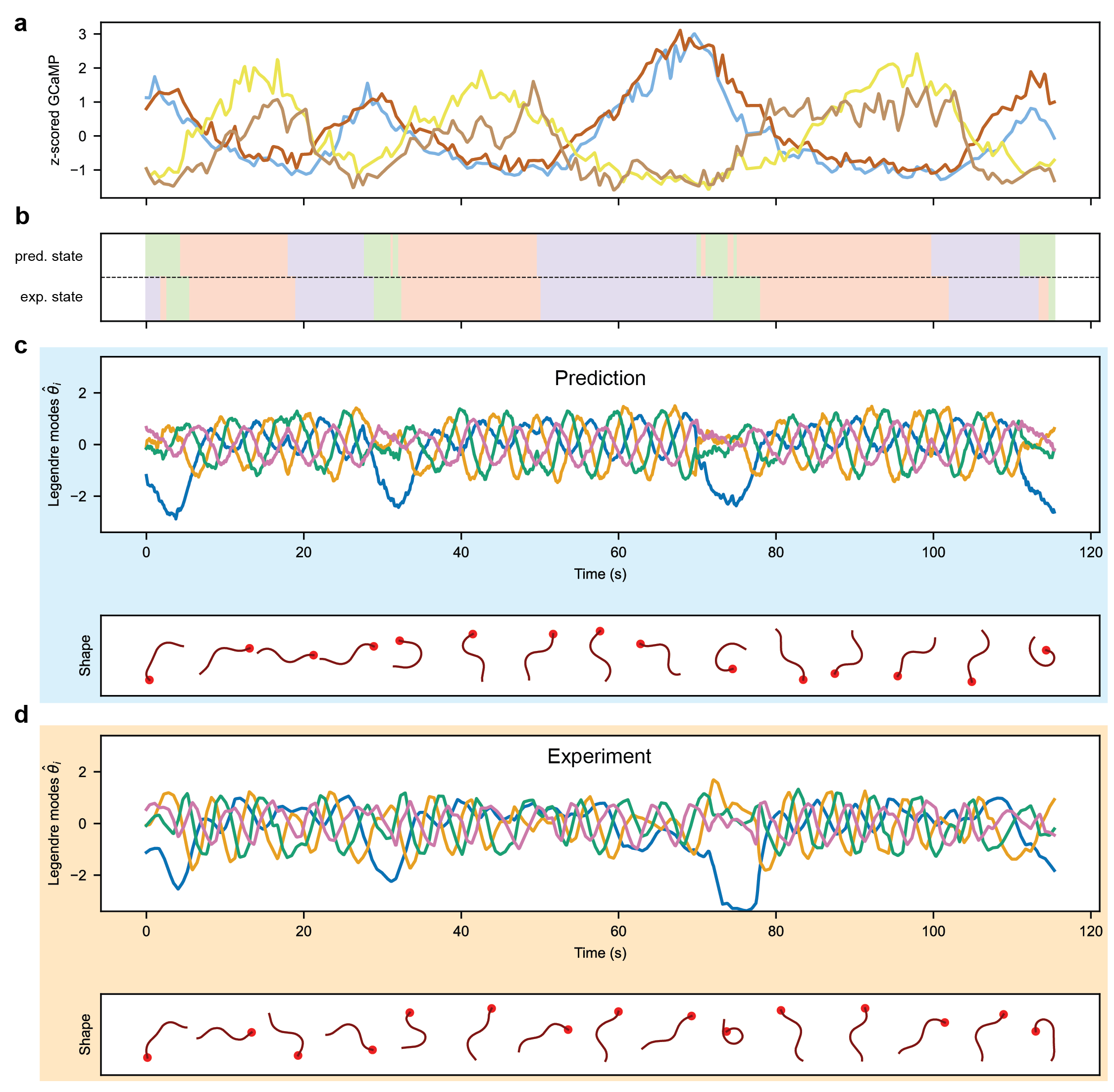}
\caption{\label{fig:fig5} 
The learned model accurately predicts typical animal postures and locomotion patterns from neural activity.
(a) Recorded neural activity from a 2-minute portion of data not included in the training set. 
(b) Comparison between a representative behavioral sequence predicted by the stochastic model (top) from the data in (a) and experimentally observed state (b) showing broad agreement (locomotion states are colored as in Figs.~\ref{fig:fig2}d and~\ref{fig:fig3}a).
(c) Predicted posture dynamics using the full nonlinear model with the behavioral trajectory inferred from the neural/behavioral model. Posture snapshots from the predicted dynamics, with rotational motion predicted from linear resistive force theory, at time points equally spaced along the 2 minute trajectory.
(d) Experimentally observed posture dynamics during the same 2-minute interval as the neural data. Posture snapshots from the experimental data at time points equally spaced along the 2-minute trajectory.} 
\end{figure*}

\textbf{Predicting motion from neural activity}.
To validate the above  framework, we use our model to generate shape modes from neural activity, and thus predict animal motion dynamics from neural recordings. 
By performing posterior inference on a neural trace withheld from the training data (Fig.~\ref{fig:fig5}a), we obtain an estimate of the most likely behavioral state sequence, which closely matches the behavioral state sequence computed directly from the mode dynamics in this test set (Fig.~\ref{fig:fig5}b).
This inferred behavioral state sequence is then fed into our SDE-MS behavioral model to generate predictions for the shape mode dynamics (Fig.~\ref{fig:fig5}c). 
The shape modes are translated into rotational body motion with resistive force theory (Fig.~\ref{fig:fig5}c, Supplementary Video 5).
Although our multi-scale model is stochastic and hence cannot match experimental  motion dynamics exactly, its predictions for behavioral state-transitions and shape mode  trajectories agree well with experimentally observed time-series data~(Fig.~\ref{fig:fig5}d).

\begin{figure*}[t]
\includegraphics[]{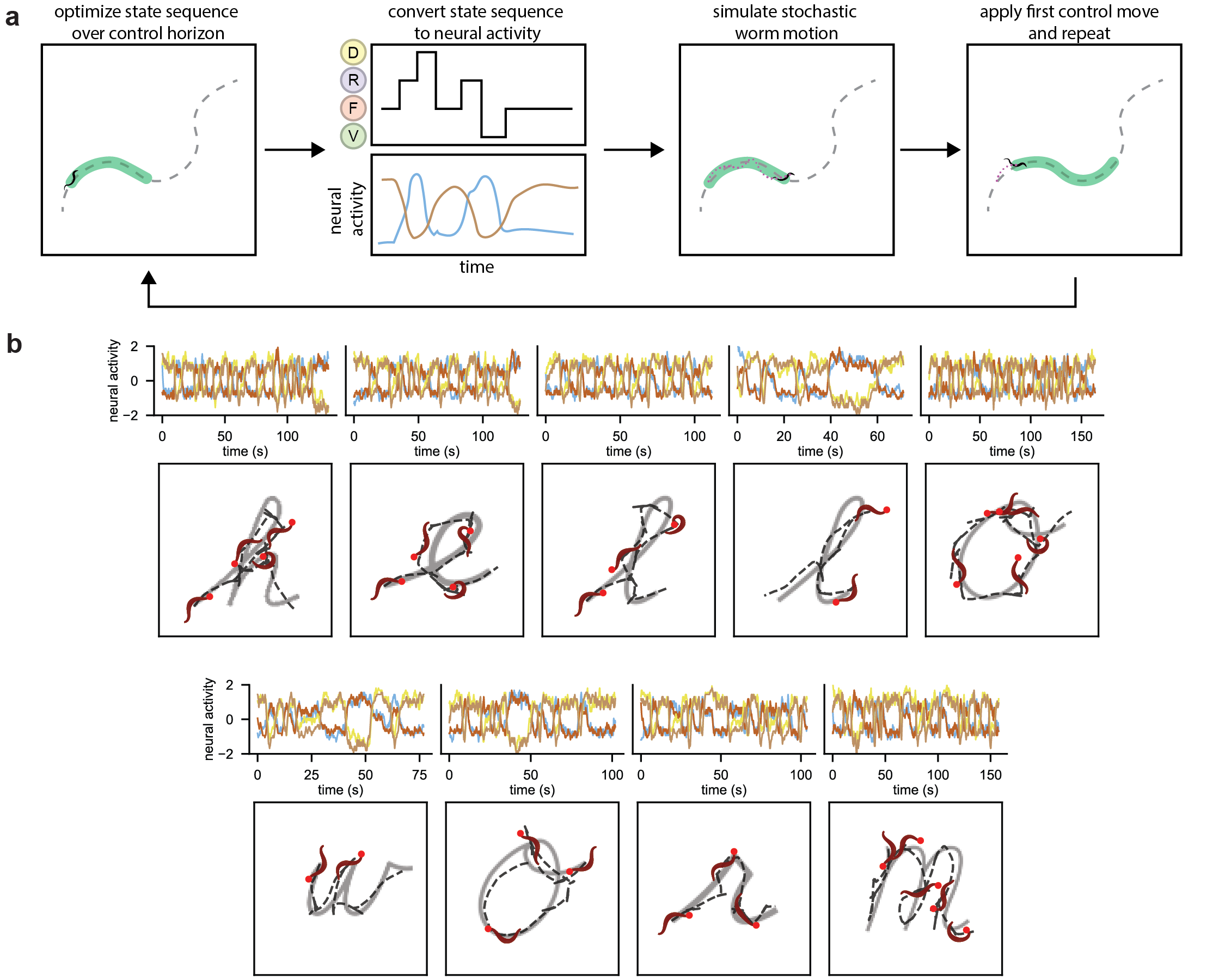}
\caption{\label{fig:fig6} 
Stochastic model  control enables the prediction of neural activity patterns that can steer worm locomotion along an arbitrary predefined path. 
(a) A finite-length segment beginning at the worm's current position is selected from the user-defined  path. Optimization finds the best-fit behavioral-state sequence to drive the worm along the path, which is then converted to a neural time series. Worm position is stochastically advanced by the first optimized control move and the control segment is updated before the process repeats.
(b) Predicted neural activity traces to steer worm motion along various target paths (gray letters). Neural control traces are shown above each trajectory, and the traces look qualitatively similar to experimentally recorded traces of the same neurons. Worm center of mass trace is indicated by the dashed black line and the worm posture is plotted in maroon every 40 seconds.} 
\end{figure*}

\textbf{Real-time control of locomotion through neural activity}.
Equally importantly, the learned model enables the prediction of neural activity given shape. To  demonstrate this and provide guidance for future optogenetic experiments, we implemented a stochastic model predictive control (MPC) framework that allows to guide worms in real-time method along a desired trajectory by inducing suitable activity patterns in selected neurons  (Fig.~\ref{fig:fig6}a). More specifically, our MPC method optimizes the behavioral state sequence such that the worm's center of mass most closely follows a finite-length segment of the complete path.
From this optimized sequence, we compute the most likely neural activity traces.
We then simulate the stochastic worm motion given these neural inputs for one control step, advance the control horizon forward, and repeat this process iteratively until the worm reaches the end of the path (see Supplementary Section XIV for a detailed description of the MPC algorithm).
To demonstrate the effectiveness of this control strategy, we direct the stochastic worm trajectory to trace several predefined target patterns (Fig.~\ref{fig:fig6}b, Supplementary Video 6).

\section*{Conclusions}
By combining spectral mode representations with a decomposition of stochastic dynamical systems into gradient and curl components, we introduced a data-efficient framework for inferring a neuromechanical locomotion model for \textit{C. elegans}. 
The approach is well-suited for current experimental constraints and data limitations, as the results presented here were achieved using fewer than 1000 time points, representing less than 10 minutes of recordings. 
The nonlinear gradient dynamics of the inferred SDE model guide the system onto the appropriate shape manifold for a given behavior, whereas a linear curl component suffices to generate motion along this manifold.
Formulating curl dynamics through generalized Nambu Hamiltonians provides a direct  geometric interpretation of the shape dynamics,  enabling efficient model parameterizations. 
The resulting multi-scale model successfully generates postures for each behavioral state and predicts motion from neural activity in agreement with experiment (Figs.~\ref{fig:fig5}), and it can be used to predict neural activity patterns that guide the worm along desired trajectories (Fig.~\ref{fig:fig6}).
Our framework could be readily adapted to model neuromechanical behavior in other animals, including those requiring more complex shape representations beyond centerlines, by using mode representations based on the geometric or graph structure of the animal's appendages~\cite{gosztolai2021liftpose3d}.
The same Hamiltonian modeling strategy can also be applied to more complex neural representations, as illustrated for mouse hippocampus data in Supplementary Section XV. 
Given the generic nature and physical interpretability of our approach at both the representation and dynamical systems levels, multi-scale models obtained within this framework can facilitate comparisons between different individuals of the same species, mutant variants, and distinct species, as well as guide optogenetic experiments that aim to control behavior through the targeted manipulation of neural activity.

\subsection*{Acknowledgements}

We thank Steven Flavell and his lab members for sharing and explaining the experimental data. We thank Tosif Ahamed for bringing reference~\cite{nambu1973generalized} to our attention, and Alasdair Hastewell, George Stepaniants, Andre Souza, Sreeparna Pradhan, and Steven Flavell for numerous helpful discussions. This research received support through Schmidt Sciences, LLC (to J.D.), and the MathWorks Professorship Fund (to J.D.). A.E.C. was funded through a National Defense Science and Engineering Graduate Fellowship. 

\clearpage
\bibliography{Ref.bib}

\clearpage

\end{document}


\title{Supplementary Information}
\maketitle

\centering{\bf Authors}: Alexander E. Cohen, Jörn Dunkel

\justifying

\vspace{1em}

\section{Data processing}
\textbf{Experimental methods}. 
The dataset used in this study was obtained from a recent publication~\cite{atanas2023brain}, which introduced a method for simultaneously recording body motion and neural activity in freely moving \textit{C. elegans}.
In their experiments, the authors employed a spinning disk confocal microscope to measure GCaMP fluorescence, while a separate optical path captured near-infrared (NIR) bright-field images of the worm’s body.
To identify individual neurons in the GCaMP recordings, they used NeuroPAL imaging in immobilized worms, allowing them to map neuron identities to the dynamic recordings.
In our work, we used the centerline extracted from the NIR images and the z-scored neural activity traces in the analysis.
For more information on the original experimental techniques, consult the original paper.
An animation of the data used in this paper can be found in Supplementary Video 1.

\textbf{Preprocessing}.
Due to limitations of the worm centerline extraction algorithm used in~\cite{atanas2023brain}, some frames contain physically unrealistic worm body postures. 
Thus, we first manually identified frames with these physically unrealistic body postures using a program we wrote to efficiently label worm postures.
We also manually identified frames where the head and tail positions are flipped, which often occurs after omega turns.
While this process can be automated, it only takes $\approx 5$ minutes to perform manually on our dataset.
After this labeling, physically unrealistic frames are removed and flipped frames are corrected.

\begin{figure*}[t]
\includegraphics[width = 1.0\textwidth]{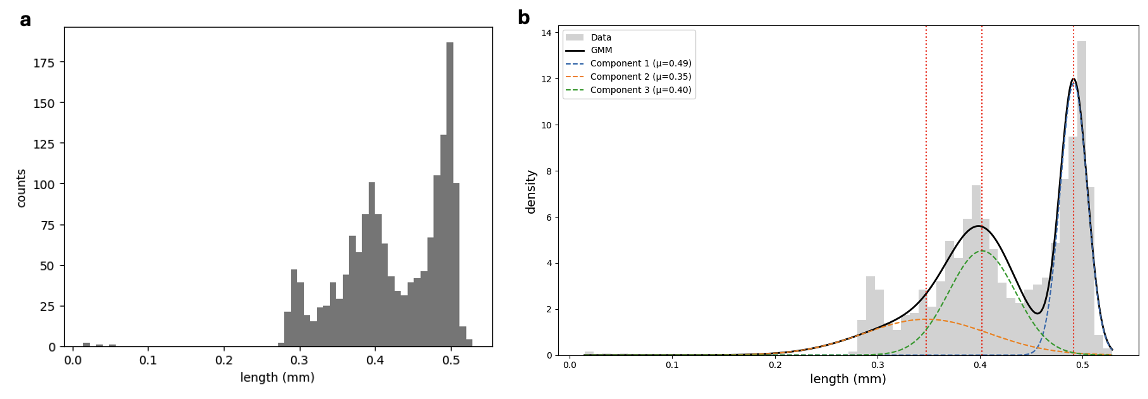}\textit{}
\caption{\label{fig:SI_length_distribution} 
(a) Distribution of tracked worm's length from the NIR field of view during the experiment.
(b) Gaussian Mixture Model fit to the distribution to identify the worm length.
} 
\end{figure*}

Additional preprocessing is necessary to account for the limited field of view of the NIR microscope.
The worm used in this paper is $\approx 0.5$ mm long.
However, there are many frames where part of the worm's length is occluded from the NIR field of view (Supp. Fig.~\ref{fig:SI_length_distribution}a). 
To mitigate this occlusion effect on our dataset, we crop the worm's length to $60\%$ of its full length, chosen because most frames include $60\%$ of the worm's body posture (Supp. Fig.~\ref{fig:SI_length_distribution}a). 
To perform the cropping, we first estimate the worm's actual length by fitting a Gaussian Mixture Model (GMM) to the distribution of lengths computed from each frame (Supp. Fig.~\ref{fig:SI_length_distribution}b).
Since the distribution shows three clear peaks, we fit a three-component GMM and use the largest mean of the three Gaussians as the worm's length (Supp. Fig.~\ref{fig:SI_length_distribution}b). 
Then, we define a cutoff at $0.6$ times this length and crop all frames to this cutoff.
If the worm's length does not meet this cutoff, we exclude the frame.
This preprocessing step ensures all frames contain the same crop of the total worm body. 
We note that this will impact the accuracy of the resistive force theory predictions, but we do not model this effect. 
To overcome this, future work can develop a model to predict the remaining 40\% of the worm's body from the first 60\%.

To get data at equally spaced time points, we linearly interpolate the worm's shape and neural activity.
We note that linear interpolation theoretically does not preserve the statistical properties of the time series data, but we have also tried higher-order interpolation, Brownian bridges, and Gaussian process regression for interpolation, and the final results and conclusions are unchanged.
In addition, we found that performing interpolation in tangent angle space $\theta(s,t)$, as opposed to real $(x(s,t),y(s,t))$ space, produces more realistic interpolated worm shapes. 

After converting the worm's shape in $(x,y)$ space to the tangent angle as a function of distance from head to tail, $\theta(s)$, we compute the Legendre modes, $\boldsymbol{\hat{\theta}}$, which are used as our low-dimensional representation of worm shape throughout the paper.
The Legendre coefficients are computed using a least-squares fit. 

\section{Shape representation}
\begin{figure*}[t]
\includegraphics[width = 1.0\textwidth]{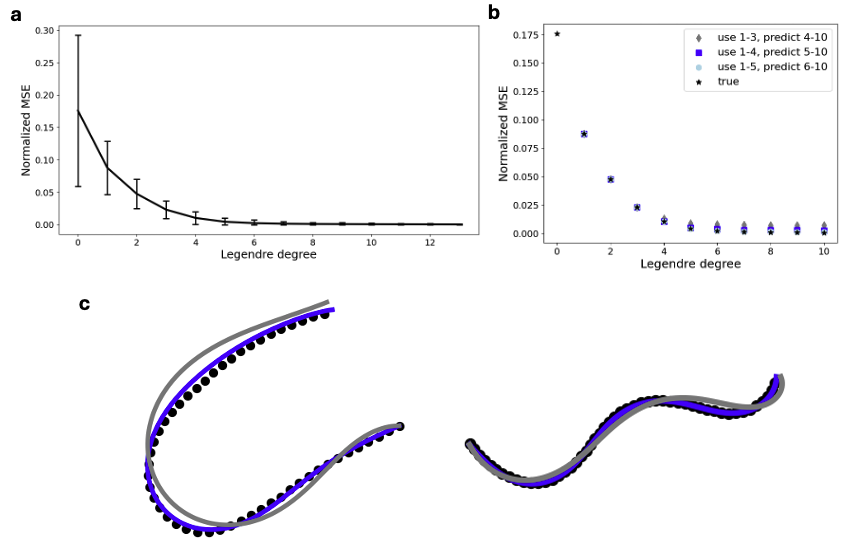}\textit{}
\caption{\label{fig:reconstruction} 
(a) Reconstruction error versus maximum Legendre polynomial degree plateaus around degree 7-8.
(b) Superresolution using a random forest regression model can accurately predict higher Legendre coefficients from lower coefficients, suggesting we only need to model up to degree 4 Legendre coefficients.
(c) Reconstructed worm shapes by using degree 1-3 coefficients plus superresolution model (grey) and using degree 1-4 coefficients plus superresolution model (blue), compared with true worm shape (black).
} 
\end{figure*}

One hyperparameter of our model is the number of Legendre modes to include in our posture representation. 
While we use orthogonal polynomials for the worm posture~\cite{cohen2023schrodinger}, other animals can be similarly represented with an orthogonal basis~\cite{romeo2021learning}.
For example, for animals with appendages, one option is to use the eigenvectors of the graph Laplacian of the connectivity matrix for the animal's joints~\cite{curreli2025nonisotropic}. 
For any representation, the choice of how many modes to keep is an important hyperparameter.

To determine the minimal dimensions of our shape representation, we examine the reconstruction error decay with increasing number of modes. 
The Legendre polynomial reconstruction error begins to plateau after approximately a degree 7-8 polynomial reconstruction (Supp. Fig.~\ref{fig:reconstruction}a). 
However, we hypothesized that we could achieve similar reconstruction errors with fewer Legendre modes by training a super-resolution model to predict higher-degree Legendre coefficients from the lower-degree coefficients.
In fact, a random forest regression model trained to predict Legendre mode degrees 5-10 from 1-4 achieves reconstruction accuracy comparable to the original Legendre mode degrees 1-10 (Supp. Fig.~\ref{fig:reconstruction}b).
Other regression techniques were also tested, including linear regression and neural networks, but the random forest model achieved the highest validation accuracy.
In addition, the actual reconstructed worm shapes when using degrees 1-4 to predict 5-10 look significantly closer to the true worm shape than when using degrees 1-3 to predict 4-10 (Supp. Fig.~\ref{fig:reconstruction}c).
Thus, we opt to only include Legendre modes 1-4 for the subsequent dynamical analysis, and we use the random forest model to predict degrees 5-10 when converting to real space worm posture from the Legendre mode representation.
(Note: we do not include the degree 0 coefficient since this encodes the worm's global orientation, which is predicted from the higher modes with resistive force theory~\cite{keaveny2017predicting,costa2024markovian}).

\section{SDE with Markov Switching and Autoregressive Hidden Markov Models}

Throughout our work, we use a stochastic differential equation with Markov switching to represent the behavioral-motion model:
\begin{equation}d\boldsymbol{\hat{\theta}} = \mathbf{f}_z(\boldsymbol{\hat{\theta}})\,dt + \boldsymbol{\Sigma}^{1/2}_z\,d\boldsymbol{W}(t)
\end{equation}

The hidden state $z$ evolves according to a continuous-time Markov chain (CTMC), with transition probabilities~\cite{aldridge2024math2750}
\begin{equation}
    \text{Pr}(z(t+h) = j \mid z(t) = i) = \delta_{ij} + q_{i,j}h + o(h).
\end{equation}
where $\delta_{ij}$ is the Kronecker delta symbol and $\mathbf{Q}$ is the transition rate matrix with entries $q_{i,j}$.

More generally, we can think of these behavioral states as the positions of the minima of an energy function on some $d$-dimensional energy landscape~\cite{pearce2019learning}.
We can then approximate a Brownian motion by computing the Kramers transition rates and using them as the generator in the continuous-time Markov chain.
An advantage of this framework is that behavior is represented as a continuous variable on an underlying landscape, rather than as intrinsically discrete states. This allows for smooth interpolation between behavioral modes and yields a more coherent dynamical model. Extending our approach in this direction is a promising avenue for future work.

To simulate and perform parameter estimation and inference on this system, we first discretize the equations in time~\cite{yuan2004convergence, nguyen2019euler}.
The CTMC is simulated using a discrete-time Markov chain with a transition matrix equal to 
\begin{equation}
    \mathbf{P}^{\Delta t} = e^{\Delta t Q}.
    \label{eq:discrete_markov}
\end{equation}

Using this discrete-time Markov chain, we discretize the SDE via the Euler-Maruyama method:
\begin{equation}
    \boldsymbol{\hat{\theta}}(t + \Delta t) = \boldsymbol{\hat{\theta}}(t) + \mathbf{f}_{z(t)}(\boldsymbol{\hat{\theta}}(t))\Delta t + \boldsymbol{\Sigma}^{1/2}_{z(t)}\Delta \mathbf{W}(t),
    \label{eq:euler_mayurama}
\end{equation}
where $\Delta \boldsymbol{W}(t)$ are independent and identically distributed random variables with mean zero and variance $\Delta t$.
For linear $\mathbf{f}_z$, simulating this system is mathematically equivalent to a specific class of probabilistic state space models called autoregressive hidden Markov Models (ARHMMs), which have been used throughout neuroscience and animal behavior~\cite{wiltschko2015mapping,lee2024switching}. 

To make this more clear, we note that the observation model for an ARHMM looks like
\begin{equation}
    p(\mathbf{x}_t \mid \mathbf{x}_{t-1}, z_t=j) = \mathcal{N}(\mathbf{f}_j(\mathbf{x}_{t-1}),\boldsymbol{\Sigma}_j),
\end{equation}
where for an ARHMM $\mathbf{f}_j$ is linear such that $\mathbf{f}_j(\mathbf{x}) = \mathbf{A}_j\mathbf{x} + \mathbf{b}_j$.
In addition, the hidden state transition model for an ARHMM is 
\begin{equation}
    p(z_t = j \mid z_{t-1} = i) = P_{ij}.
\end{equation}
The full generative model just combines the observation and transition model to give
\begin{equation}
    p(\mathbf{x}_{1:T}, z_{1:T}) = \bigg[p(z_1)\prod_{t=2}^Tp(z_t \mid z_{t-1})\bigg]\bigg[p(\mathbf{x}_1 \mid z_1)\prod_{t=2}^T p(\mathbf{x}_t \mid \mathbf{x}_{t-1}, z_t)\bigg].
\end{equation}

Similarly, we can rearrange Equations \eqref{eq:discrete_markov} and \eqref{eq:euler_mayurama} to write our approximate time-discretized SDE-MS model as
\begin{equation}
    p(\boldsymbol{\hat{\theta}}(t+\Delta t) \mid  \boldsymbol{\hat{\theta}}(t), z(t) = j) = \mathcal{N}(\boldsymbol{\hat{\theta}}(t) + \mathbf{f}_j(\boldsymbol{\hat{\theta}}(t))\Delta t, \boldsymbol{\Sigma}_j \Delta t)
\end{equation}
and 
\begin{equation}
    p(z(t+\Delta t)=j \mid z(t)=i) = \mathbf{P}^{\Delta t}_{ij}
    \end{equation}

\section{Stationary Distribution}

For an SDE (without Markov switching), $\dot{\mathbf{x}} = \mathbf{f}(\mathbf{x}) + \boldsymbol{\Sigma}^{1/2}\boldsymbol{\xi}$, where $\boldsymbol{\Sigma}$ is a constant symmetric matrix, we can write a partial differential equation (PDE) that describes the evolution of the probability density of the state, $\rho(\mathbf{x},t)$, called the Fokker-Planck equation, as
\begin{equation}
    \frac{\partial \rho}{\partial t} = -\nabla \cdot \bigg(\mathbf{f} \rho - \frac{1}{2}\boldsymbol{\Sigma}\nabla\rho\bigg)
    \label{eq:fokker_planck}
\end{equation}
To solve for the stationary distribution, we set $\frac{\partial\rho}{\partial t} = 0$. 
Let us assume that the vector field takes the form 
\newline
$\mathbf{f} = -\frac{1}{2} \boldsymbol{\Sigma}\nabla \Psi + \mathbf{g}$.
We will now show that the steady-state distribution has the Boltzmann form $\rho \propto \exp{(-\Psi)}$ if \mbox{$\nabla \cdot (\mathbf{g}\exp{(-\Psi)})=0$}.
After plugging in these expressions for $\mathbf{f}$ and $\rho$ into Equation~\eqref{eq:fokker_planck}, we obtain
\begin{equation}
    \nabla \cdot \bigg[\bigg(-\frac{1}{2}\boldsymbol{\Sigma}\nabla \Psi + \mathbf{g}\bigg)\exp{(-\Psi)} + \frac{1}{2}\boldsymbol{\Sigma}\nabla \Psi\exp{(-\Psi)}\bigg]=0.
\end{equation}
After canceling the $\pm \frac{1}{2}\boldsymbol{\Sigma}\nabla \Psi$ terms, we are left with 
\begin{equation}
    \nabla \cdot (\mathbf{g} \exp{(-\Psi)}) = 0
\end{equation}
which can be broken up using the product rule and then simplified to:
\begin{equation}
    \nabla(-\Psi)\cdot \mathbf{g} + \nabla\cdot\mathbf{g} = 0.
    \label{eq:general_constraint}
\end{equation}

While Eq.~\eqref{eq:general_constraint} represents the general constraint such that $\mathbf{g}$ does not influence the stationary distribution, in our work we apply a more restrictive version where both terms in the sum are set approximately or exactly to zero. 
In particular, we satisfy $\nabla \Psi \cdot \mathbf{g}=0$ approximately with data augmentation and $\nabla\cdot\mathbf{g}=0$ exactly by constraining $\mathbf{g}$ to be a divergence-free field.

We note that the above derivation is true for a single SDE, without Markov switching. 
Therefore, we individually apply this method to each SDE within our SDE-MS. 
In the main text, we use this construction for the state-conditioned stationary distributions of the individual SDEs, while representing the observed marginal distribution over behavioral states as an occupancy-weighted mixture.

\section{Helmholtz decomposition in diffusion-whitened coordinates}

In this section we show that the decomposition
\begin{equation}
\mathbf{f}(\mathbf{x})
=
-\frac{1}{2}\boldsymbol{\Sigma}\nabla_{\mathbf{x}}\Psi(\mathbf{x})
+
\mathbf{g}(\mathbf{x})
\label{eq:decomposition_final}
\end{equation}
arises naturally by performing a standard Helmholtz decomposition in coordinates where the diffusion tensor is isotropic and then transforming back to the original variables.

\textbf{Whitening the SDE}. 
Consider the It\^o SDE
\begin{equation}
d\mathbf{x}_t
=
\mathbf{f}(\mathbf{x}_t)\,dt
+
\boldsymbol{\Sigma}^{1/2} d\mathbf{W}_t,
\end{equation}
where $\boldsymbol{\Sigma}$ is constant, symmetric, and positive definite.
Define the linear change of variables
\begin{equation}
\mathbf{y}
=
\boldsymbol{\Sigma}^{-1/2}\mathbf{x}.
\end{equation}
Because this transformation is linear, It\^o's formula introduces no additional drift correction terms. The SDE becomes
\begin{equation}
d\mathbf{y}_t
=
\tilde{\mathbf{f}}(\mathbf{y}_t)\,dt
+
d\mathbf{W}_t,
\end{equation}
where
\begin{equation}
\tilde{\mathbf{f}}(\mathbf{y})
=
\boldsymbol{\Sigma}^{-1/2}
\mathbf{f}(\boldsymbol{\Sigma}^{1/2}\mathbf{y}).
\end{equation}
In the $\mathbf{y}$ variables the diffusion tensor is the identity.

\subsection*{Standard Helmholtz decomposition in whitened coordinates}

With isotropic diffusion, the deterministic drift can be decomposed using the standard Euclidean Helmholtz decomposition:
\begin{equation}
\tilde{\mathbf{f}}(\mathbf{y})
=
-\frac{1}{2}
\nabla_{\mathbf{y}}\tilde{\Psi}(\mathbf{y})
+
\tilde{\mathbf{g}}(\mathbf{y}),
\qquad
\nabla_{\mathbf{y}}\cdot\tilde{\mathbf{g}}(\mathbf{y})=0.
\label{eq:helmholtz_y}
\end{equation}
The first term is a gradient field and the second term is divergence-free.

\subsection*{Transforming back to the original variables}

We now express this decomposition in terms of $\mathbf{x}$.
From the definition of $\tilde{\mathbf{f}}$,
\begin{equation}
\mathbf{f}(\mathbf{x})
=
\boldsymbol{\Sigma}^{1/2}
\tilde{\mathbf{f}}(\boldsymbol{\Sigma}^{-1/2}\mathbf{x}).
\end{equation}
Substituting Eq.~\eqref{eq:helmholtz_y} gives
\begin{align}
\mathbf{f}(\mathbf{x})
&=
\boldsymbol{\Sigma}^{1/2}
\left(
-\frac{1}{2}
\nabla_{\mathbf{y}}\tilde{\Psi}
+
\tilde{\mathbf{g}}
\right)
\Bigg|_{\mathbf{y}=\boldsymbol{\Sigma}^{-1/2}\mathbf{x}}.
\end{align}
Define the scalar potential in $\mathbf{x}$-coordinates by
\begin{equation}
\Psi(\mathbf{x})
=
\tilde{\Psi}(\boldsymbol{\Sigma}^{-1/2}\mathbf{x}).
\end{equation}
Using the chain rule,
\begin{equation}
\nabla_{\mathbf{x}}\Psi
=
\boldsymbol{\Sigma}^{-1/2}
\nabla_{\mathbf{y}}\tilde{\Psi},
\qquad
\text{or equivalently}
\qquad
\nabla_{\mathbf{y}}\tilde{\Psi}
=
\boldsymbol{\Sigma}^{1/2}
\nabla_{\mathbf{x}}\Psi.
\end{equation}
Substituting this relation yields
\begin{align}
\mathbf{f}(\mathbf{x})
&=
-\frac{1}{2}
\boldsymbol{\Sigma}^{1/2}
\left(
\boldsymbol{\Sigma}^{1/2}
\nabla_{\mathbf{x}}\Psi
\right)
+
\boldsymbol{\Sigma}^{1/2}
\tilde{\mathbf{g}}(\boldsymbol{\Sigma}^{-1/2}\mathbf{x}) \\
&=
-\frac{1}{2}
\boldsymbol{\Sigma}
\nabla_{\mathbf{x}}\Psi
+
\mathbf{g}(\mathbf{x}),
\end{align}
where
\begin{equation}
\mathbf{g}(\mathbf{x})
=
\boldsymbol{\Sigma}^{1/2}
\tilde{\mathbf{g}}(\boldsymbol{\Sigma}^{-1/2}\mathbf{x}).
\end{equation}
This recovers exactly the form in Eq.~\eqref{eq:decomposition_final}.

\subsection*{Preservation of divergence-free and orthogonality conditions}

We now show that both divergence-freeness and the orthogonality condition are preserved under the whitening transformation
\[
\mathbf{y}=\boldsymbol{\Sigma}^{-1/2}\mathbf{x},
\qquad
\mathbf{x}=\boldsymbol{\Sigma}^{1/2}\mathbf{y}.
\]
Recall that the curl component transforms as
\begin{equation}
\mathbf{g}(\mathbf{x})
=
\boldsymbol{\Sigma}^{1/2}
\tilde{\mathbf{g}}(\mathbf{y}),
\qquad
\mathbf{y}=\boldsymbol{\Sigma}^{-1/2}\mathbf{x},
\label{eq:g_transform_clean}
\end{equation}
and that the potentials are related by
\begin{equation}
\Psi(\mathbf{x})
=
\tilde{\Psi}(\mathbf{y})
=
\tilde{\Psi}(\boldsymbol{\Sigma}^{-1/2}\mathbf{x}).
\label{eq:psi_transform_clean}
\end{equation}

\textbf{Divergence-free condition.}
Let $J_{\mathbf{x}}\mathbf{g}$ denote the Jacobian matrix of $\mathbf{g}$ with respect to $\mathbf{x}$, and similarly $J_{\mathbf{y}}\tilde{\mathbf{g}}$ the Jacobian of $\tilde{\mathbf{g}}$ with respect to $\mathbf{y}$.
From Eq.~\eqref{eq:g_transform_clean}, applying the chain rule for Jacobians gives
\begin{equation}
J_{\mathbf{x}}\mathbf{g}
=
\boldsymbol{\Sigma}^{1/2}
\,
\left(J_{\mathbf{y}}\tilde{\mathbf{g}}\right)
\,
\boldsymbol{\Sigma}^{-1/2}.
\end{equation}
The divergence is the trace of the Jacobian:
\[
\nabla_{\mathbf{x}}\cdot\mathbf{g}
=
\mathrm{tr}(J_{\mathbf{x}}\mathbf{g}).
\]
Therefore,
\begin{align}
\nabla_{\mathbf{x}}\cdot\mathbf{g}
&=
\mathrm{tr}
\!\left(
\boldsymbol{\Sigma}^{1/2}
(J_{\mathbf{y}}\tilde{\mathbf{g}})
\boldsymbol{\Sigma}^{-1/2}
\right).
\end{align}
Using cyclicity of the trace, $\mathrm{tr}(ABC)=\mathrm{tr}(BCA)$, we obtain
\begin{equation}
\nabla_{\mathbf{x}}\cdot\mathbf{g}
=
\mathrm{tr}
\!\left(
(J_{\mathbf{y}}\tilde{\mathbf{g}})
\boldsymbol{\Sigma}^{-1/2}
\boldsymbol{\Sigma}^{1/2}
\right)
=
\mathrm{tr}(J_{\mathbf{y}}\tilde{\mathbf{g}})
=
\nabla_{\mathbf{y}}\cdot\tilde{\mathbf{g}}.
\end{equation}
Hence,
\begin{equation}
\nabla_{\mathbf{x}}\cdot\mathbf{g}=0
\quad\Longleftrightarrow\quad
\nabla_{\mathbf{y}}\cdot\tilde{\mathbf{g}}=0.
\end{equation}

\textbf{Orthogonality condition.}
From Eq.~\eqref{eq:psi_transform_clean}, the chain rule for gradients under a linear change of variables gives
\begin{equation}
\nabla_{\mathbf{x}}\Psi
=
\boldsymbol{\Sigma}^{-1/2}
\nabla_{\mathbf{y}}\tilde{\Psi}.
\label{eq:grad_relation_clean}
\end{equation}
Using Eqs.~\eqref{eq:g_transform_clean} and \eqref{eq:grad_relation_clean},
\begin{align}
\nabla_{\mathbf{x}}\Psi \cdot \mathbf{g}
&=
\left(
\boldsymbol{\Sigma}^{-1/2}
\nabla_{\mathbf{y}}\tilde{\Psi}
\right)
\cdot
\left(
\boldsymbol{\Sigma}^{1/2}
\tilde{\mathbf{g}}
\right) \\
&=
(\nabla_{\mathbf{y}}\tilde{\Psi})^{T}
\left(
\boldsymbol{\Sigma}^{-1/2}
\boldsymbol{\Sigma}^{1/2}
\right)
\tilde{\mathbf{g}} \\
&=
(\nabla_{\mathbf{y}}\tilde{\Psi})^{T}
\tilde{\mathbf{g}}
=
\nabla_{\mathbf{y}}\tilde{\Psi}
\cdot
\tilde{\mathbf{g}}.
\end{align}
Therefore,
\begin{equation}
\nabla_{\mathbf{x}}\Psi\cdot\mathbf{g}=0
\quad\Longleftrightarrow\quad
\nabla_{\mathbf{y}}\tilde{\Psi}\cdot\tilde{\mathbf{g}}=0.
\end{equation}

\medskip

Thus, by whitening the SDE, performing a standard Helmholtz decomposition, and transforming back to the original coordinates, we obtain precisely the decomposition used in the main text, which separates the components responsible for the stationary distribution ($\boldsymbol{\Sigma}$-weighted gradient) from the components responsible for the nonequilibrium flow (divergence-free $\mathbf{g}$).

\subsection*{Coordinate convention for Hamiltonian surfaces}

The whitening construction above is used to motivate the form of the decomposition in Eq.~\eqref{eq:decomposition_final}. In the model fits and figures, however, the Nambu Hamiltonians are parameterized after transforming back to the original variables $\mathbf{x}$, which for the worm data are the Legendre-mode coordinates. Thus, unless explicitly denoted with tildes, the Hamiltonians and their ellipsoid-plane intersections are surfaces in the original data coordinates, not in the diffusion-whitened coordinates.

If one instead starts with Hamiltonians $\tilde H_i(\mathbf{y})$ in whitened coordinates, then the corresponding level sets in the original variables are obtained by the pullback
\begin{equation}
H_i(\mathbf{x})=\tilde H_i(\boldsymbol{\Sigma}^{-1/2}\mathbf{x}).
\end{equation}
The transformed field $\mathbf{g}(\mathbf{x})=\boldsymbol{\Sigma}^{1/2}\tilde{\mathbf{g}}(\boldsymbol{\Sigma}^{-1/2}\mathbf{x})$ remains a Nambu-Hamiltonian curl field in $\mathbf{x}$ coordinates. Indeed, the Nambu vector field generated by the pulled-back Hamiltonians is proportional to $\mathbf{g}(\mathbf{x})$ by the constant factor $\det(\boldsymbol{\Sigma}^{-1/2})$, which can be absorbed by rescaling one Hamiltonian. Thus, because the whitening transformation is linear and state-independent, the divergence-free and Hamiltonian properties of the curl component are preserved under the transformation. Consequently, the geometric objects shown for the learned worm model should be interpreted as Hamiltonian surfaces for the curl component in the original Legendre-mode coordinates. Equivalent surfaces in whitened coordinates would be related by the linear map $\mathbf{y}=\boldsymbol{\Sigma}^{-1/2}\mathbf{x}$.

\section{Denoising Score Matching}
To parameterize and learn $\Psi$, we use a machine learning method known as denoising score matching. 
The main goal behind score matching is to learn the gradient of the log of a probability distribution (the data distribution), also known as the score function, $\mathbf{s}(\mathbf{x}) = \nabla_\mathbf{x}\log{(p(\mathbf{x}))}$.
This has many applications in generative modeling, most popularly in diffusion models which are commonly trained with a score matching loss function.

To perform score matching, we parameterize a score function $\mathbf{s}_\theta$ and optimize it to match the true score function of the data distribution.
We can write this optimization problem with the objective function
\begin{equation}
    \mathbb{E}_{\mathbf{x} \sim p_{\text{data}}(\mathbf{x})}\bigg[\bigg|\bigg| \nabla_\mathbf{x}\log{p(\mathbf{x})} - \mathbf{s}_\theta(\mathbf{x}) \bigg|\bigg|^2_2\bigg].
\end{equation}
However, this objective function is impossible to optimize because we do not have access to the true data distribution or score function of the data distribution.
Instead, there is an equivalent loss function that only differs by a constant and thus has the same minimizer.
This loss function is known as denoising score matching:
\begin{equation}
    \mathbb{E}_{\mathbf{x} \sim p_{\text{data}}(\mathbf{x}), \tilde{\mathbf{x}} \sim q_\sigma (\tilde{\mathbf{x}} \mid \mathbf{x})} \bigg[\bigg | \bigg | \nabla_{\tilde{\mathbf{x}}} \log{q(\tilde{\mathbf{x}} \mid \mathbf{x})} - \mathbf{s}_\theta(\tilde{\mathbf{x}}) \bigg | \bigg |^2_2\bigg],
\end{equation}
where $q_\sigma (\tilde{\mathbf{x}} \mid \mathbf{x})$ is a noise distribution usually taken to be Gaussian, which simplifies the objective function to
\begin{equation}
    \mathbb{E}_{\mathbf{x} \sim p_{\text{data}}(\mathbf{x}), \tilde{\mathbf{x}} \sim \mathcal{N}(\tilde{\mathbf{x}}; \mathbf{x}, \sigma^2)} \bigg[\bigg|\bigg| \frac{1}{\sigma^2}(\mathbf{x} - \tilde{\mathbf{x}}) - \mathbf{s}_\theta(\tilde{\mathbf{x}}) \bigg|\bigg|^2_2\bigg].
\end{equation}

\begin{figure*}[t]
\includegraphics[width = 0.3\textwidth]{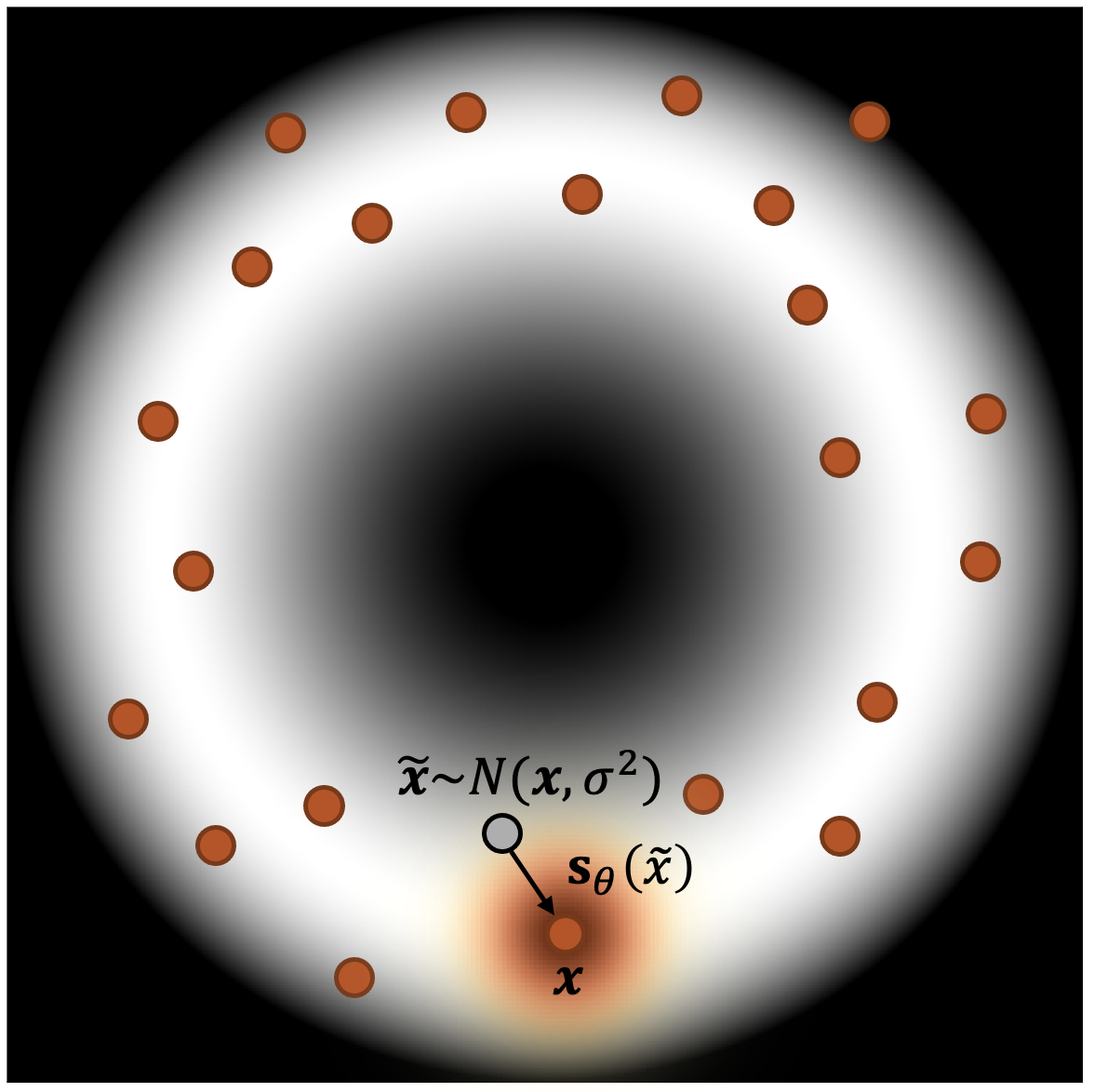}\textit{}
\caption{\label{fig:SI_DSM} 
Visual illustration of denoising score matching. 
The background color indicates the true data distribution, where whiter color indicates higher probability regions. 
The orange points represent the data points, which are samples from the true data distribution.
Denoising score matching effectively adds noise to the data points according to $\tilde{\mathbf{x}} \sim \mathcal{N}(\mathbf{x}, \sigma^2)$ and predicts how to reverse this noise with the score function $\mathbf{s}_\theta$.
The data approximately live on a circular manifold, and the score function points towards this manifold.
} 
\end{figure*}

Intuitively, this formulation of the problem is called denoising score matching because you are essentially adding noise to the data and training a score function to undo or ``denoise'' the effect of the noise (Supp. Fig.~\ref{fig:SI_DSM}).
If you hypothesize that your data live on a low-dimensional manifold, you can interpret the score function as moving you closer to the data manifold.

Alternatively, one could also learn the gradient component of the dynamics by training a diffusion model to generate data from the stationary distribution~\cite{arts2023two}.
The forward process of a diffusion model consists of gradually adding more noise to the data (with a Markov process or an SDE, although they are equivalent).
In order to generate data, we must reverse or denoise this process, which involves the use of a score model $\mathbf{s}_{\boldsymbol{\theta}}(\mathbf{x}_i,i)$, where $i$ indicates the step in the Markov chain.
In theory, this score model will equal the score of the marginal distribution of $\mathbf{x}_i$ at the $i$th step/noise level.
At low noise levels/steps, this distribution is nearly the data distribution, so the score function here will approximate the score of the true data distribution.
Thus, in practice one can train a diffusion model to generate data from the stationary distribution and then use the score function evaluated at a low noise level/step as the gradient component of the dynamics.
We use this approach in our analysis of the mouse hippocampus data.

\section{Nambu Hamiltonian Dynamics}
In order to parameterize our curl, or divergence-free, dynamics, we use a generalization of Hamiltonian dynamics to higher dimensional phase spaces~\cite{nambu1973generalized}.
For an $n$-dimensional state $\mathbf{x}$, we can express generalized Hamiltonian dynamics using $n-1$ Hamiltonians $H_i$ as
\begin{equation}
    \frac{dx_i}{dt} = \sum_{j,k,\ldots,l} \epsilon_{ijk\ldots l} \frac{\partial H_1}{\partial x_j} \frac{\partial H_2}{\partial x_k} \cdots \frac{\partial H_{n-1}}{\partial x_l}.
    \label{eq:hamiltonian_dynamics}
\end{equation}
which is divergence-free~\cite{nambu1973generalized}.

Furthermore, these dynamics are determined by the intersection of the surfaces defined by $H_i = c_i$, where the $c_i$ are constants set by the initial conditions.
This feature provides a convenient geometric interpretation of the dynamics, which in practice can be used to place constraints on the Hamiltonians.
In our worm case, in each state the shape modes tend to oscillate with one dominant frequency (see wavelet analysis below), so we constrain the Hamiltonians, in the original Legendre-mode coordinates, to ellipsoids and planes.
For the mouse hippocampus data, the CEBRA embeddings live on a sphere, so we constrain one of the Hamiltonians to be this sphere~\cite{schneider2023learnable}.
This parameterization is equivalent to a vector potential parameterization, but provides the additional benefits of being geometrically interpretable and enabling geometric constraints~\cite{axenides2010strange}.

A nice demonstration of Nambu Hamiltonian dynamics examines the Lorenz system~\cite{axenides2010strange}.
The Lorenz equations can be broken up into irrotational and rotational components, 
\begin{equation}
    \frac{d}{dt}
    \begin{pmatrix}
        x \\ 
        y \\
        z
    \end{pmatrix} = 
    \begin{pmatrix}
        -\sigma x \\ 
        -y \\
        -\beta z
    \end{pmatrix} +
    \begin{pmatrix}
        \sigma y \\ 
        x(\rho-z) \\
        xy
    \end{pmatrix}.
    \label{eq:lorenz}
\end{equation}
The first term in Eq.~\eqref{eq:lorenz} is the irrotational component, which comes from the gradient of the potential function $V(x,y,z) = -\frac{1}{2}(\sigma x^2 + y^2 + \beta z^2)$. 
Level sets of this potential function are plotted in Supplementary Figure~\ref{fig:nambu_lorenz}a.
The second term in Eq.~\eqref{eq:lorenz} is the rotational component, which comes from the two generalized Hamiltonians, $H_1 = \frac{1}{2}(y^2 + (z-\rho)^2)$ and $H_2 = \sigma z - \frac{x^2}{2}$.
$H_1$ is a cylinder that extends along the $x$ axis and is centered at $(y,z)=(0,\rho)$ while $H_2$ is a parabola extending along the $y$ axis.
The intersection of these surfaces for specific values of the parameters $\rho$ and $\sigma$ approximates the butterfly wing structure of the Lorenz attractor~(Supp. Fig.~\ref{fig:nambu_lorenz}b).
The combination of these surfaces and the potential produce the Lorenz dynamics (Supp. Fig.~\ref{fig:nambu_lorenz}). 

\begin{figure*}[t]
\includegraphics[]{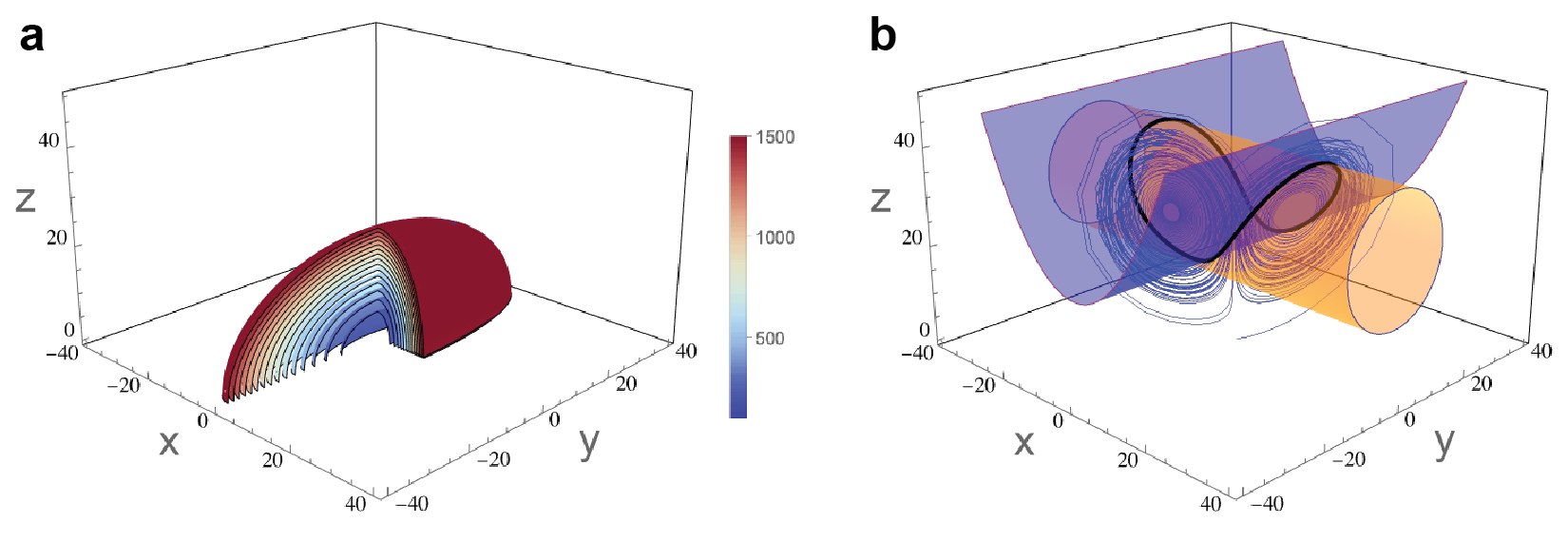}
\caption{\label{fig:nambu_lorenz} 
Geometric Hamiltonian representation of the Lorenz equations. 
(A) Irrotational part of the Lorenz dynamics illustrated as a contour plot of the function whose gradient produces the irrotational dynamics.
(B) The two Hamiltonian surfaces (blue and orange) that produce the rotational dynamics that lie on the intersection of the surfaces (black). An example Lorenz trajectory is overlaid (blue line), which appears near the black surface intersection.
} 
\end{figure*}

\section{Simplification and factorization of distribution}

As discussed in the main text, the goal is to learn a model that captures the distribution of neural activity, motion, and behavior, which can be expressed as
\begin{equation}
    p(z(t)=i,\boldsymbol{\hat{\theta}}(t),\mathbf{n}(t) \mid z(t')=j,\boldsymbol{\hat{\theta}}(t'),\mathbf{n}(t'))
	    \label{eq:full_distribution}
\end{equation}
for $t'<t$.
Given that we approximate the fully observed system to be Markovian, if we discretize in time we seek to model the distribution for $t' = t - \Delta t$.

Experimentally, there exist only a very small number of datasets that contain neural activity, motion, and behavior, making it difficult to directly infer a model for Eq.~\eqref{eq:full_distribution}~\cite{atanas2023brain}.
Even within the datasets that do include neural activity, behavior, and motion, the neural recordings only contain a subset of all neurons of interest.
However, there are significantly more motion recordings than neural recordings because the technology for obtaining these measurements is older and the experiments are easier to run~\cite{yemini2013database}.
The sampling frequencies of the microscopes used for recording worm motion are also much higher than the microscopes used for recording neural activity due to the fact that imaging neural activity requires 3D scans through the worm's body.
Specifically, in the dataset we use, the motion data is sampled at $40$ Hz compared with $1.4$--$1.7$ Hz for the neural data~\cite{atanas2023brain}.

Taken together, these facts suggest a factorization of the distribution that both simplifies the process of learning a model and makes the model more useful from an experimental perspective. 
This factorization involves splitting the model into a neural component for relating neural activity to motion and behavior, and a behavioral component for modeling motion and behavior.
Mathematically, this split can be expressed as
\begin{equation}
    p(z(t)=i,\boldsymbol{\hat{\theta}}(t),\mathbf{n}(t) \mid z(t')=j,\boldsymbol{\hat{\theta}}(t'),\mathbf{n}(t')) = p(\mathbf{n}(t) \mid \boldsymbol{\hat{\theta}}(t), z(t)=i, \mathbf{n}(t')) p(\boldsymbol{\hat{\theta}}(t), z(t)=i \mid \boldsymbol{\hat{\theta}}(t'), z(t')=j),
\end{equation}
where we refer to the first term on the right-hand side as the neural distribution and the second term as the behavioral distribution.

We can approximate the behavioral distribution to first order with an Euler-Maruyama approximation given by $p(\boldsymbol{\hat{\theta}}(t), z(t)=i \mid \boldsymbol{\hat{\theta}}(t'), z(t')=j) = (\delta_{ij}+q_{j,i}h)\mathcal{N}(\boldsymbol{\hat{\theta}}(t')+\mathbf{f}_{j}(\boldsymbol{\hat{\theta}}(t'))h, \boldsymbol{\Sigma}_{j} h)$,
where $h=t-t'$, which is accurate for small $h$.
For the neural distribution, we still have the problem that each experiment only records a subset of the neurons of interest.
To mitigate this partial observability, we can leverage a conditional independence assumption demonstrated in~\cite{atanas2023brain}.
From~\cite{atanas2023brain}, the authors built a model to predict neural activity from specific behavioral variables.
In particular, they model each neuron independently, and demonstrate that you can predict individual neural activity accurately given behavior alone (not including other neurons). 
Motivated by this work, we can simplify the neural model using conditional independence of neurons given behavior as,
\begin{equation}
    p(\mathbf{n}(t) \mid \mathbf{n}(t'), z(t),\boldsymbol{\hat{\theta}}(t)) = \prod_{i=1}^N p(n_i(t) \mid n_i(t'), z(t), \boldsymbol{\hat{\theta}}(t)).
    \label{eq:neural_dist_simplified}
\end{equation}
For this work, since we focus on neurons that encode forward, reverse, and turning motions, we also drop posture from the right-hand side.
Furthermore, rather than including $n_i(t')$ as a given variable, we model the distribution of neural activity and its time derivative,
\begin{equation}
    \prod_{i=1}^N p(n_i(t), \dot{n}_i(t) \mid z(t)).
    \label{eq:neural_model}
\end{equation}

\section{State identification}
Identifying the latent states is a critical part of model inference and parameter estimation.
In previous work using ARHMMs~\cite{wiltschko2015mapping}, this is done by using the model directly with Bayesian methods.
First, parameter estimation can be performed with expectation-maximization (EM) algorithms or with gradient descent to minimize the negative log likelihood of the observed states~\cite{murphy2022probabilistic}.
Given a set of parameters, the Viterbi algorithm can be used to compute the most likely hidden (behavioral) states at each time point~\cite{murphy2022probabilistic}.
In our work, to account for the Nambu Hamiltonian parameterization of the curl dynamics, we use maximum likelihood/gradient descent methods instead of EM methods.
Complex parameterization can easily be incorporated into the gradient-based methods using bijector functions, while some of the closed-form parameter updates in the EM updates may need to be rederived in what can be a complicated procedure~\cite{Linderman_Dynamax_A_Python_2025}.
This standard approach worked well for the example system in Fig.~1, but performed worse for the experimental data.
We believe that the main reason for this is the limited amount of experimental data, specifically the limited amount of dorsal and ventral turns in the dataset.
While the model accurately identifies and captures the forward and backward motion, it fails to reliably identify the turning states.
However, this is not surprising because there are only $\approx 9$ dorsal turns and $\approx 25$ ventral turns in the entire dataset.
There are a similar number of tracking failures, which cause unphysical spikes in the mode time dynamics, so the model is unable to differentiate between real motion states and noisy artifacts in the data.

Therefore, we instead develop a method that first clusters the time series based on local geometric features of the dynamics~\cite{cohen2023schrodinger,costa2019adaptive}.
This method involves fitting Nambu Hamiltonians in the original Legendre-mode coordinates to short time segments ($\approx 2$ seconds) of data and clustering on simple geometric features of the Hamiltonian surfaces. 
In particular, we use the orientation of the surfaces, as this specifies the direction the state moves along the intersection, and the maximum value of the intersection in the curvature Legendre mode dimension.
Supplementary Movie 2 shows how the surface orientations and intersections of the Nambu Hamiltonians fit to $2$-second segments of the data change in time over the course of the dataset.

\subsection{Wavelet-based state discovery}
While not used in the main results, we also explored the use of wavelet transforms to identify behavioral states, similar to~\cite{berman2014mapping}.
Wavelet transforms can be used to visualize distinct dynamical regimes in the animal's motion.
Commonly used in meteorology and acoustics, wavelet analysis allows us to extract how dominant dynamical modes vary in time~\cite{torrence1998practical}. 

Mathematically, animal motion often exhibits oscillatory patterns on short time scales during movement gaits, such as the undulations of worms and snakes or the steps of legs in insects and quadrupedal and bipedal animals.
Over longer timescales, the dynamics of these oscillations change as the animal moves at different speeds, changes direction, and displays distinct behavioral states, each characterized by different stereotypical postural sequences~\cite{cermak2020whole, costa2024markovian}.

\begin{figure*}[t]
\includegraphics[width = 1.0\textwidth]{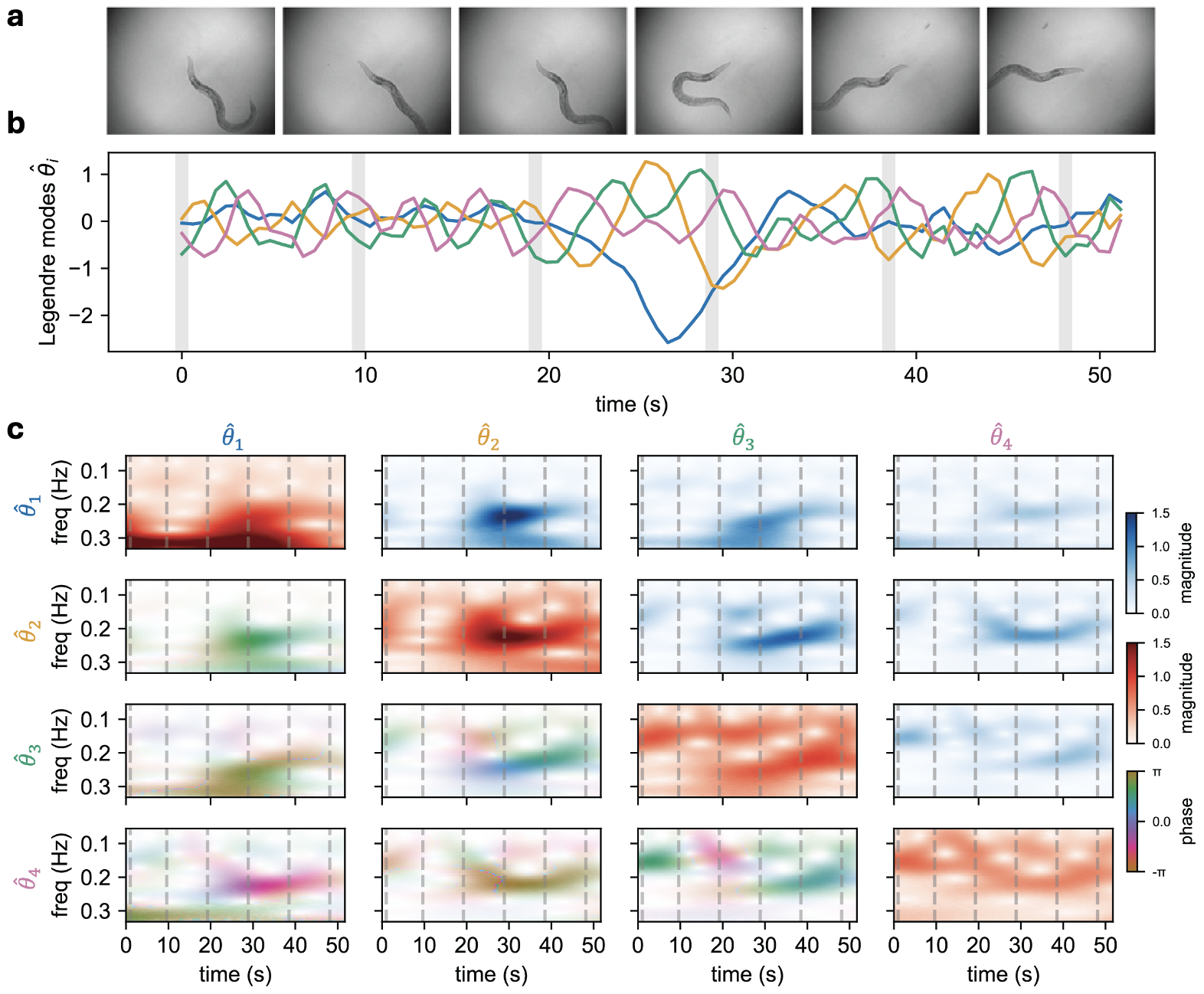}
\caption{\label{fig:wavelets} 
Wavelet scalograms and coherence distinguish behavioral states in worm motion.
Representative trajectory of the Legendre representation of worm shape (a) and the corresponding wavelet scalograms and coherences (b). 
The coherence phase undergoes a sign flip upon the transition between a forward and reverse state. 
} 
\end{figure*}

The time series data, along with wavelet scalograms and coherences, distinguish the distinct forward, reverse, and turning states (Supp. Fig.~\ref{fig:wavelets}).
When the worm performs a large bend before a turn, $\hat{\theta}_1$ spikes to a large negative value for ventral bends and a large positive value for dorsal bends (Supp. Fig.~\ref{fig:wavelets}a).
During forward and reverse motion, $\hat{\theta}_{2-4}$ oscillate at a limited number of frequencies (diagonal and upper right plots of Supp. Fig.~\ref{fig:wavelets}b).
In the forward state, the phase between $\hat{\theta}_3$ and $\hat{\theta}_4$ is positive (green), while in the reverse state the phase is negative (pink) (lower right of Supp. Fig.~\ref{fig:wavelets}).

\subsection{Spike-weighted phase–curvature distributions}

\begin{figure*}[t]
\includegraphics[width = 1.0\textwidth]{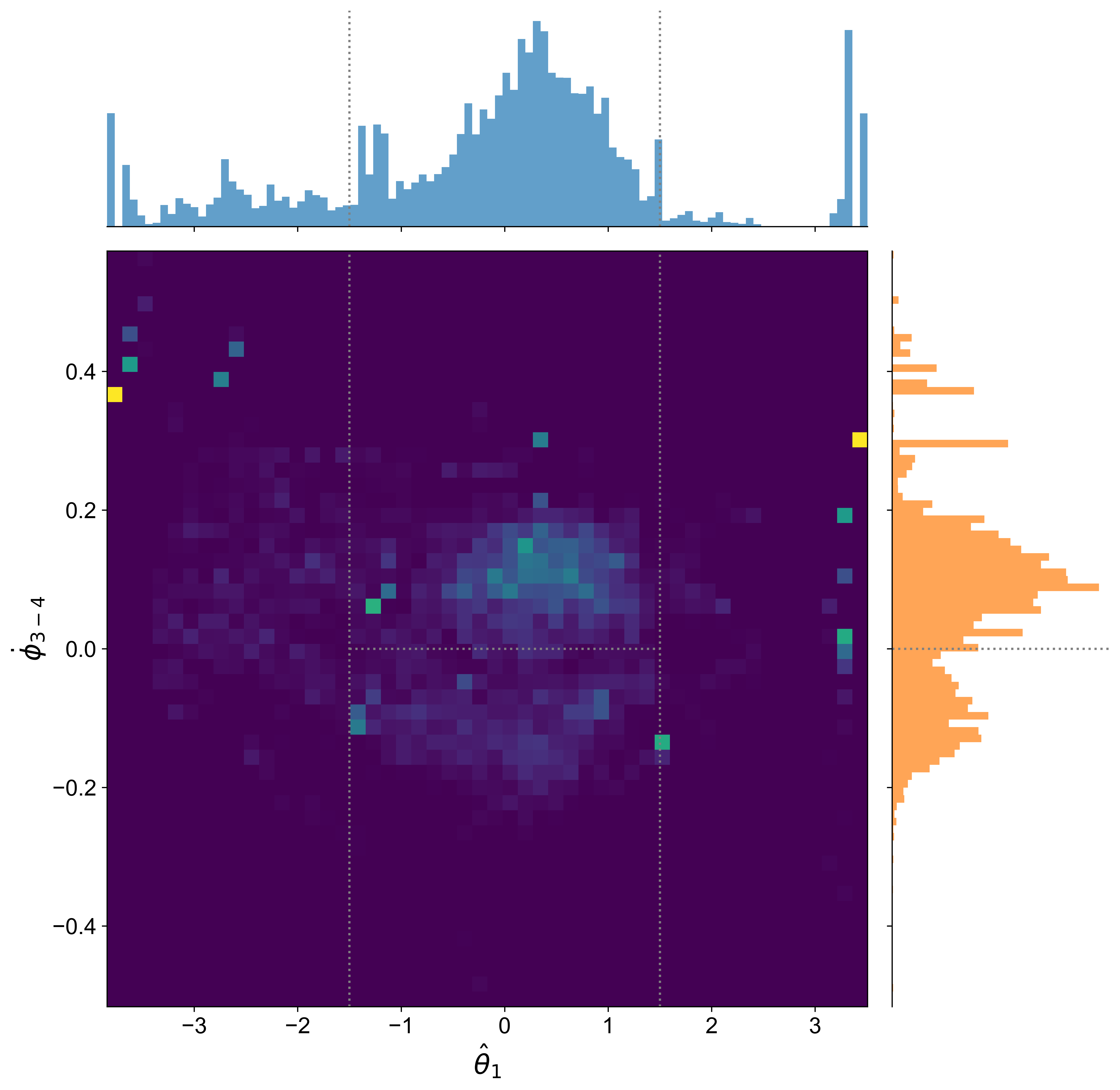}
\caption{\label{fig:2d_hist_marginals} 
Spike-weighted joint distribution of curvature and phase velocity reveals transient turning dynamics.
Two-dimensional histogram of the circular curvature Legendre mode $\hat{\theta}_1$ (x-axis) and the phase velocity between modes $\hat{\theta}_3$ and $\hat{\theta}_4$ (y-axis). 
Each time point is weighted by the magnitude of the high-pass residual of $\hat{\theta}_1$, emphasizing short-lived curvature excursions associated with turning events while retaining all observations.
Top and right panels show the corresponding weighted marginal distributions.
Dashed lines indicate reference regions used to distinguish dynamical regimes.
} 
\end{figure*}

Because turning events are rare in the experimental dataset, standard time-averaged statistics obscure the dynamical signatures associated with these behaviors. 
In particular, large dorsal and ventral turns appear as short-lived excursions in the Legendre mode dynamics and therefore contribute negligibly to conventional histograms or density estimates that weight all time points equally.
To better visualize dynamical regimes associated with brief but behaviorally relevant events, we construct spike-weighted joint distributions of geometric observables.
Specifically, we compute a high-pass residual of the circular curvature mode,
\begin{equation}
r(t) = \hat{\theta}_1(t) - \tilde{\theta}_1(t),
\end{equation}
where $\tilde{\theta}_1(t)$ is obtained by Gaussian smoothing of $\hat{\theta}_1(t)$ over timescales longer than typical undulatory motion. The magnitude of this residual,
\begin{equation}
w(t) = |r(t)|,
\end{equation}
serves as a measure of local geometric transience and assigns larger statistical weight to short-lived curvature excursions associated with turning behavior.

We then compute a weighted two-dimensional histogram between the circular curvature mode $\hat{\theta}_1$ and the phase velocity between Legendre modes $\hat{\theta}_3$ and $\hat{\theta}_4$ (Fig.~\ref{fig:2d_hist_marginals}). 
Each time point contributes weight $w(t)$ to the corresponding histogram bin, yielding a spike-conditioned density. 
This procedure emphasizes regions of state space visited during rapid geometric transitions while preserving the full dataset.
Marginal distributions shown alongside the joint histogram correspond to the same spike-weighted statistics (Fig.~\ref{fig:2d_hist_marginals}).

\section{Behavioral Model Inference Procedure}
The full inference procedure for our model consists of steps to learn both the divergence-free component of the model and the gradient component.
The data tend to be dominated by the divergence-free component.
This is because, as discussed before, the gradient component serves to push the data onto the correct behavioral distribution/manifold of postures, while the divergence-free component specifies the dynamics on this distribution/manifold.
Therefore, most of the data lie on maxima of the probability distribution, where the gradient or score of the distribution is zero or very small.

We start by learning the divergence-free component of the dynamics.
To do this, we need a parameterization of the dynamics in terms of parameters of the Hamiltonians.
Since we use four Legendre coefficients, as motivated above, we need three Hamiltonian functions to represent the divergence-free dynamics.
Within each state, there are one or two dominant frequencies of oscillations of the modes (discussed above and shown in Supplementary Fig.~\ref{fig:wavelets}).
Therefore, we can use the geometric interpretation of the Hamiltonian dynamics to constrain the dynamics within a set that captures these simple oscillations.
All Hamiltonians in this inference procedure are written in the original Legendre-mode coordinates, $\mathbf{x}=\hat{\boldsymbol{\theta}}$.
In this case, the intersection between an ellipsoid and two planes produces single-frequency oscillations.
Using these shapes as the Hamiltonians also constrains the divergence-free dynamics to be linear (the gradient component will still be nonlinear).

We parameterize the ellipsoid through the Cholesky decomposition of a positive definite matrix, as the equation for an ellipsoid can be written as $\mathbf{x}^\top\mathbf{A}\mathbf{x}=1$, where $\mathbf{A}$ is positive definite.
We parameterize each plane with the equation $H_p = p_1 \hat{\theta}_1 + p_2 \hat{\theta}_2 + p_3 \hat{\theta}_3 + p_4 \hat{\theta}_4$.
Since this set of dynamics is linear, we can then derive the dynamics matrix by evaluating Eq.~\eqref{eq:hamiltonian_dynamics}.
Specifically, for Hamiltonians:
\begin{align*}
    H_1 &= a_{11}x_1^2 + 2a_{12}x_1x_2 + 2a_{13}x_1x_3 + 2a_{14}x_1x_4 + a_{22}x_2^2 + 2a_{23}x_2x_3 + 2a_{24}x_2x_4 + a_{33}x_3^2 + 2a_{34}x_3x_4 + a_{44}x_4^2 \\
    H_2 &= b_1x_1 + b_2x_2 + b_3x_3 + b_4x_4 \\
    H_3 &= c_1x_1 + c_2x_2 + c_3x_3 + c_4x_4,
\end{align*}
we derive the following dynamics matrix:
\begin{equation}
    \frac{d}{dt} \hat{\boldsymbol{\theta}} = 2 \mathbf{M}\hat{\boldsymbol{\theta}},
\end{equation}
where
\begin{align*}
    m_{11} &= -a_{14}b_3c_2 + a_{13}b_4c_2 + a_{14}b_2c_3 - a_{12}b_4c_3 - a_{13}b_2c_4 + a_{12}b_3c_4 \\
    m_{12} &= -a_{24}b_3c_2 + a_{23}b_4c_2 + a_{24}b_2c_3 - a_{22}b_4c_3 - a_{23}b_2c_4 + a_{22}b_3c_4 \\
    m_{13} &= -a_{34}b_3c_2 + a_{33}b_4c_2 + a_{34}b_2c_3 - a_{23}b_4c_3 - a_{33}b_2c_4 + a_{23}b_3c_4 \\
    m_{14} &= -a_{44}b_3c_2 + a_{34}b_4c_2 + a_{44}b_2c_3 - a_{24}b_4c_3 - a_{34}b_2c_4 + a_{24}b_3c_4 \\
    m_{21} &= a_{14}b_3c_1 - a_{13}b_4c_1 - a_{14}b_1c_3 + a_{11}b_4c_3 + a_{13}b_1c_4 - a_{11}b_3c_4 \\
    m_{22} &= a_{24}b_3c_1 - a_{23}b_4c_1 - a_{24}b_1c_3 + a_{12}b_4c_3 + a_{23}b_1c_4 - a_{12}b_3c_4 \\
    m_{23} &= a_{34}b_3c_1 - a_{33}b_4c_1 - a_{34}b_1c_3 + a_{13}b_4c_3 + a_{33}b_1c_4 - a_{13}b_3c_4 \\ 
    m_{24} &= a_{44}b_3c_1 - a_{34}b_4c_1 - a_{44}b_1c_3 + a_{14}b_4c_3 + a_{34}b_1c_4 - a_{14}b_3c_4 \\
    m_{31} &= -a_{14}b_2c_1 + a_{12}b_4c_1 + a_{14}b_1c_2 - a_{11}b_4c_2 - a_{12}b_1c_4 + a_{11}b_2c_4 \\
    m_{32} &= -a_{24}b_2c_1 + a_{22}b_4c_1 + a_{24}b_1c_2 - a_{12}b_4c_2 - a_{22}b_1c_4 + a_{12}b_2c_4 \\
    m_{33} &= -a_{34}b_2c_1 + a_{23}b_4c_1 + a_{34}b_1c_2 - a_{13}b_4c_2 - a_{23}b_1c_4 + a_{13}b_2c_4 \\
    m_{34} &= -a_{44}b_2c_1 + a_{24}b_4c_1 + a_{44}b_1c_2 - a_{14}b_4c_2 - a_{24}b_1c_4 + a_{14}b_2c_4 \\ 
    m_{41} &= a_{13}b_2c_1 - a_{12}b_3c_1 - a_{13}b_1c_2 + a_{11}b_3c_2 + a_{12}b_1c_3 - a_{11}b_2c_3 \\
    m_{42} &= a_{23}b_2c_1 - a_{22}b_3c_1 - a_{23}b_1c_2 + a_{12}b_3c_2 + a_{22}b_1c_3 - a_{12}b_2c_3 \\
    m_{43} &= a_{33}b_2c_1 - a_{23}b_3c_1 - a_{33}b_1c_2 + a_{13}b_3c_2 + a_{23}b_1c_3 - a_{13}b_2c_3 \\
    m_{44} &= a_{34}b_2c_1 - a_{24}b_3c_1 - a_{34}b_1c_2 + a_{14}b_3c_2 + a_{24}b_1c_3 - a_{14}b_2c_3
\end{align*}
We then fit these parameters and the covariance of the noise via maximum likelihood under an Euler-Maruyama discretization of the SDE.
In this procedure, we do not account for the gradient dynamics, even though they are present in the data we are using for training the curl dynamics.
However, this has little impact on the inference procedure for two reasons.
The first is that our parameterization only accounts for divergence-free terms, so it will not capture any gradient dynamics.
The second is that most of the data tend to lie near the maxima of the stationary distribution, so the gradient dynamics will tend to be near zero, as discussed above. 

After fitting the curl component, we next fit the gradient component with denoising score matching, as discussed above. 
However, we also need to be careful to satisfy the constraint of Eq.~\eqref{eq:general_constraint}.
We use data augmentation to approximately satisfy this constraint.
Here, the geometric interpretation of the generalized Hamiltonian dynamics is useful once again.
As noted earlier, the intersection of the Hamiltonian surfaces defines the dynamics. 
Therefore, this intersection should also be on the maxima of the stationary distribution, as then, at least near this intersection, the vector field lines from the curl component will point tangent to the intersection and the vector field lines from the gradient component will point perpendicular to the intersection (and this is the most important region where we want to satisfy the constraint, as the data will always live near this region).

Therefore, we augment the dataset by adding points uniformly sampled along the intersection of the Hamiltonian surfaces before using denoising score matching to fit the gradient component of the dynamics.
We find the specific level set of the Hamiltonians via a least-squares fit to the data points.
Adding these data points forces the maxima of the distribution to lie on the Hamiltonian intersection.

Before developing the above method, we planned to use Bayesian inference to learn the behavioral model without the need to precluster the posture data.
This approach is discussed in Supplementary Section IX and is how we obtained our results on the example dataset.
In this approach, we parameterize the dynamics matrices using the Nambu Hamiltonians, as described above, and then use gradient descent to minimize the negative log likelihood of the observed data, computed using forward filtering. 
We have implemented this approach in a forked version of the dynamax repository for probabilistic state space models~\cite{Linderman_Dynamax_A_Python_2025}.
This gives the divergence-free component in each behavioral state. 
Using this model, we can infer the hidden states using the Viterbi algorithm.
Finally, we can fit the gradient components with denoising score matching using the segmented data.

Other work has also explored the decomposition of dynamics into conservative and nonconservative components, with other methods for learning these components (see~\cite{giorgini2025data,pedersen2025thermalizer,sosanya2022dissipative}).

\section{Neural Model Inference Procedure}

The neural model is formulated as Eq.~\eqref{eq:neural_model}.
Mathematically, we use a categorical hidden Markov model (HMM) to model this distribution.
The activity and its time derivative of each neuron is discretized into 40 bins between $\pm 2.5$ standard deviations from its mean.
Values that fall outside this range are clipped to the maximum or minimum bin.
We then create the empirical categorical distribution for the binned values.
The neuron time series traces and corresponding categorical distributions for all neurons used in the model are shown in Supplementary Figure~\ref{fig:neural_data}.
For this categorical HMM, we use the same hidden state transition matrix that was learned for the behavioral model.

\begin{figure*}[t]
\includegraphics[width = 1.0\textwidth]{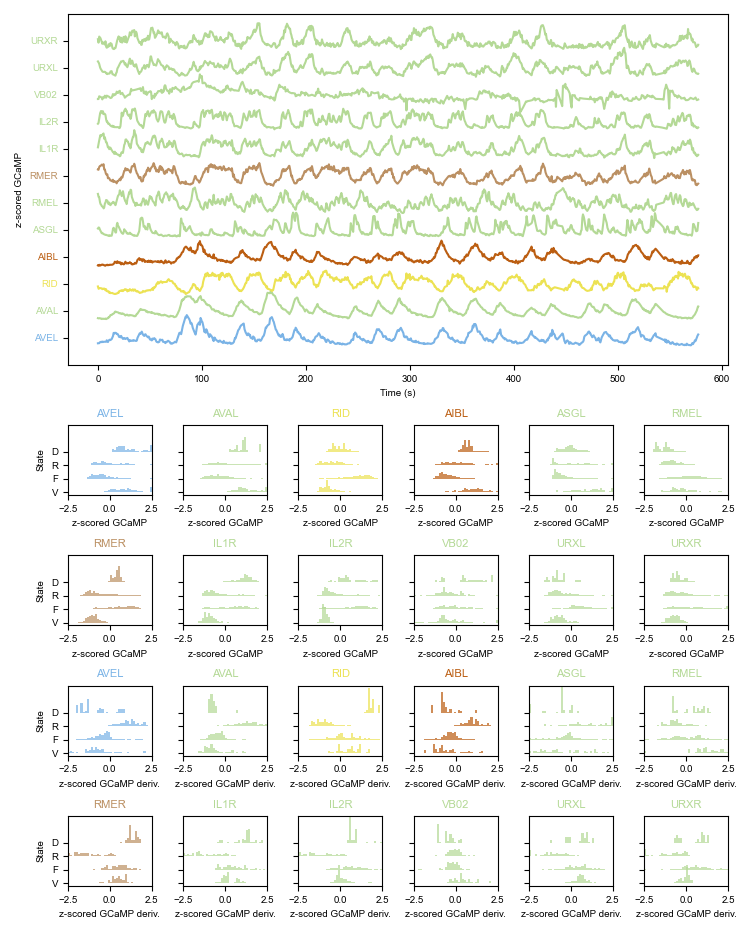}
\caption{\label{fig:neural_data} 
Time series and histograms for all neurons used in this study. 
} 
\end{figure*}

\section{Example Model}
The example system of counter-rotating elliptical limit cycles used in the main text to demonstrate our approach can be expressed as
\begin{equation}
    \frac{d}{dt}\begin{bmatrix}
        x \\
        y
    \end{bmatrix} = \mathbf{f}_z(x,y)
\end{equation}

where 
\begin{align*}
    \mathbf{f}_z = 
    &\begin{bmatrix}
    (-1)^z\frac{1}{\alpha}((x - x_0)\cos{(\theta)} - (y - y_0)\sin{(\theta)})\sin{(\theta)} - (-1)^z\alpha((x - x_0)\sin{(\theta)} + (y - y_0)\cos{(\theta)})\cos{(\theta)} \\
    (-1)^z\frac{1}{\alpha}((x - x_0)\cos{(\theta)} - (y - y_0)\sin{(\theta)})\cos{(\theta)} + (-1)^z\alpha((x - x_0)\sin{(\theta)} + (y - y_0)\cos{(\theta)})\sin{(\theta)}
    \end{bmatrix} - \\
    &\begin{bmatrix}
        (x - x_0)((\frac{1}{\alpha}((x-x_0)\cos{(\theta)} - (y-y_0)\sin{(\theta)}))^2 + ((x-x_0)\sin{(\theta)} + (y-y_0)\cos{(\theta)})^2 - 1) \\
        (y - y_0)((\frac{1}{\alpha}((x-x_0)\cos{(\theta)} - (y-y_0)\sin{(\theta)}))^2 + ((x-x_0)\sin{(\theta)} + (y-y_0)\cos{(\theta)})^2 - 1)
        \end{bmatrix}.
\end{align*}

Specifically, we set $x_0 = 1.5, y_0=1.5$ and $\theta = \frac{-\pi}{4}$ for $z=0$ and $x_0 = -1.5, y_0=1.5$ and $\theta = \frac{\pi}{4}$ for $z=1$, and $\alpha = 2.0$. 
This model can give generic elliptical limit cycles for any choices of $(x_0, y_0, \theta, \alpha)$.
These parameters set the position, orientation, and aspect ratio of the ellipses.

For simplicity, to connect this model to the Helmholtz-Nambu framework, we consider here the case where $(x_0,y_0) = (0,0)$, $\theta = 0$, and $\alpha = 1.0$. This simplifies the dynamics to
\begin{equation}
    \mathbf{f}_z = \begin{bmatrix}
    (-1)^z(-y) + x(1 - x^2 - y^2) \\
    (-1)^z x + y(1 - x^2 - y^2)
    \end{bmatrix}
    = (-1)^z\begin{pmatrix} -y \\ x \end{pmatrix} + (1 - |\mathbf{x}|^2)\begin{pmatrix} x \\ y \end{pmatrix},
    \label{eq:toy_compact}
\end{equation}
which admits an exact decomposition of the form used in Eq.~\eqref{eq:decomposition_final},
\begin{equation}
    \mathbf{f}_z = \underbrace{\mathbf{g}_z(\mathbf{x})}_{\text{curl}} + \underbrace{{}-\nabla \Psi(\mathbf{x})}_{\text{gradient}},
\end{equation}
with the Hamiltonian, curl, and potential given by
\begin{equation}
    H = \tfrac{1}{2}(x^2 + y^2), \qquad
    \mathbf{g}_z = (-1)^z\begin{pmatrix} -y \\ x \end{pmatrix}, \qquad
    \Psi = H^2 - H = \tfrac{1}{4}(x^2+y^2)^2 - \tfrac{1}{2}(x^2+y^2).
    \label{eq:toy_decomposition}
\end{equation}
The level sets of $H$ are circles, defining the orbits of the curl dynamics.
The potential $\Psi$ has a minimum at $H = \frac{1}{2}$, i.e.\ along the unit circle $|\mathbf{x}|=1$, which is exactly the stable limit cycle.
Because $\Psi$ is a function of $H$, the orthogonality condition $\nabla \Psi \cdot \mathbf{g}_z = 0$ is automatically satisfied: $\nabla\Psi = (2H-1)\nabla H$ is parallel to $\nabla H$, which is perpendicular to $\mathbf{g}_z$.
The factor $(-1)^z$ reverses the direction of circulation between the two behavioral states, while the gradient term $(1 - |\mathbf{x}|^2)\mathbf{x}$ stabilizes the amplitude in both states identically.
Notably, in this example the curl component $\mathbf{g}_z$ is exactly linear while the gradient component $-\nabla\Psi$ is nonlinear.
This structure can also be seen by linearizing the system near the limit cycle, where $|\mathbf{x}|^2 = 1$ and the gradient term vanishes, leaving only the linear curl dynamics:
\begin{equation}
    \mathbf{f}_z \big|_{|\mathbf{x}|=1} = (-1)^z\begin{pmatrix} -y \\ x \end{pmatrix}.
\end{equation}
The nonlinear gradient term $-\nabla\Psi = (1 - |\mathbf{x}|^2)\mathbf{x}$ then acts as a restoring force that stabilizes the amplitude onto this periodic orbit.
This illustrates how, even when the full dynamics is nonlinear, the curl component can remain linear (or approximately linear) with all nonlinearity confined to the gradient component. We exploit this property in our application to the \textit{C.\ elegans} and mouse data.
The general elliptical limit cycles used in Fig.~1 of the main text are obtained by reintroducing the center offsets $\mathbf{x}_0$, rotation angle $\theta$, and aspect ratio $\alpha$ as specified above.

\section{Impact of Markov behavioral state model}
\begin{figure*}[t]
\includegraphics[width = 1.0\textwidth]{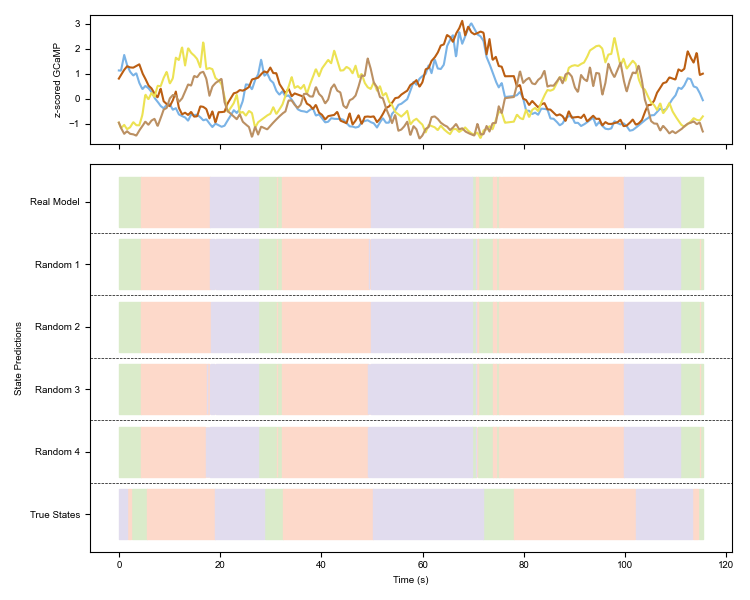}
\caption{\label{fig:markov_assumption} 
Hidden states inferred from the neural model.
(a) Neural activity.
(b) Top row shows hidden states inferred from the trained neural model. Next four rows show hidden states inferred from the trained neural model after replacing the hidden state Markov transition matrix with a randomly chosen transition matrix. Bottom row shows ``True'' hidden states as computed with our Nambu Hamiltonian local clustering on the posture mode representation time series.
} 
\end{figure*}
As discussed in the main text, the assumption that the behavioral states are Markovian is not valid in the real world, as animals have memory and the distribution of time spent in a state is not always exponential.
However, we note that this is not a major limitation of the work, since we do not use the model in the full generative sense, generating posture and neural activity and behavior. 
Instead, we use it for tasks where behavior serves as an intermediate, such as inferring shape from neural activity or optimizing neural activity given shape. 
In these cases, behavior is an intermediate between shape and neural activity, and the Markov assumption only has a small role.
To demonstrate this, we plot the behavioral state trajectories inferred from neural activity using the behavioral transition matrix learned from data and a few randomly generated transition matrices (Supp. Fig.~\ref{fig:markov_assumption}). 
The real model and the models with randomly chosen transition matrices have only minor differences in the inferred hidden states.

\section{Predicting Neural Activity to Move the Worm Along Desired Paths}
We use stochastic model predictive control to estimate the neural activity that would drive the animal along a specified trajectory.
First, we define a path along which we want the worm to travel. 
We then place the worm at the starting point of this path, oriented along the path's tangent at the starting point.
We estimate the optimal state sequence that moves the worm along approximately one worm length segment of the desired trajectory.
The state sequence is constrained such that each state lasts an integer multiple of 20 time points.
In addition, the ordering of the states follows the straight motion, reverse, turn, straight motion cycle.
These constraints ensure that the resulting dynamics appear realistic, as the worm cannot rapidly transition between different states or make unphysical transitions under these constraints.
At each optimization step, we optimize over all possible combinations of the next 5 states, or the next 100 time steps.

Once we find the optimal state sequence, we simulate the worm motion forward for the first state in the optimal sequence (20 time steps).
During this simulation, we use a different random seed than the seed we used for the optimization to achieve a more realistic setting.
The control horizon is then advanced forward by finding the closest point along the path from the updated worm position, and taking that point plus one worm's length along the desired trajectory.
This process is repeated until the worm is within a certain cutoff distance of the final point along the desired trajectory.
Finally, we obtain the predicted neural activity by simulating the neural model given the optimal state sequence.

\section{Mouse Hippocampus Data}
\begin{figure*}[t]
\includegraphics[width = 1.0\textwidth]{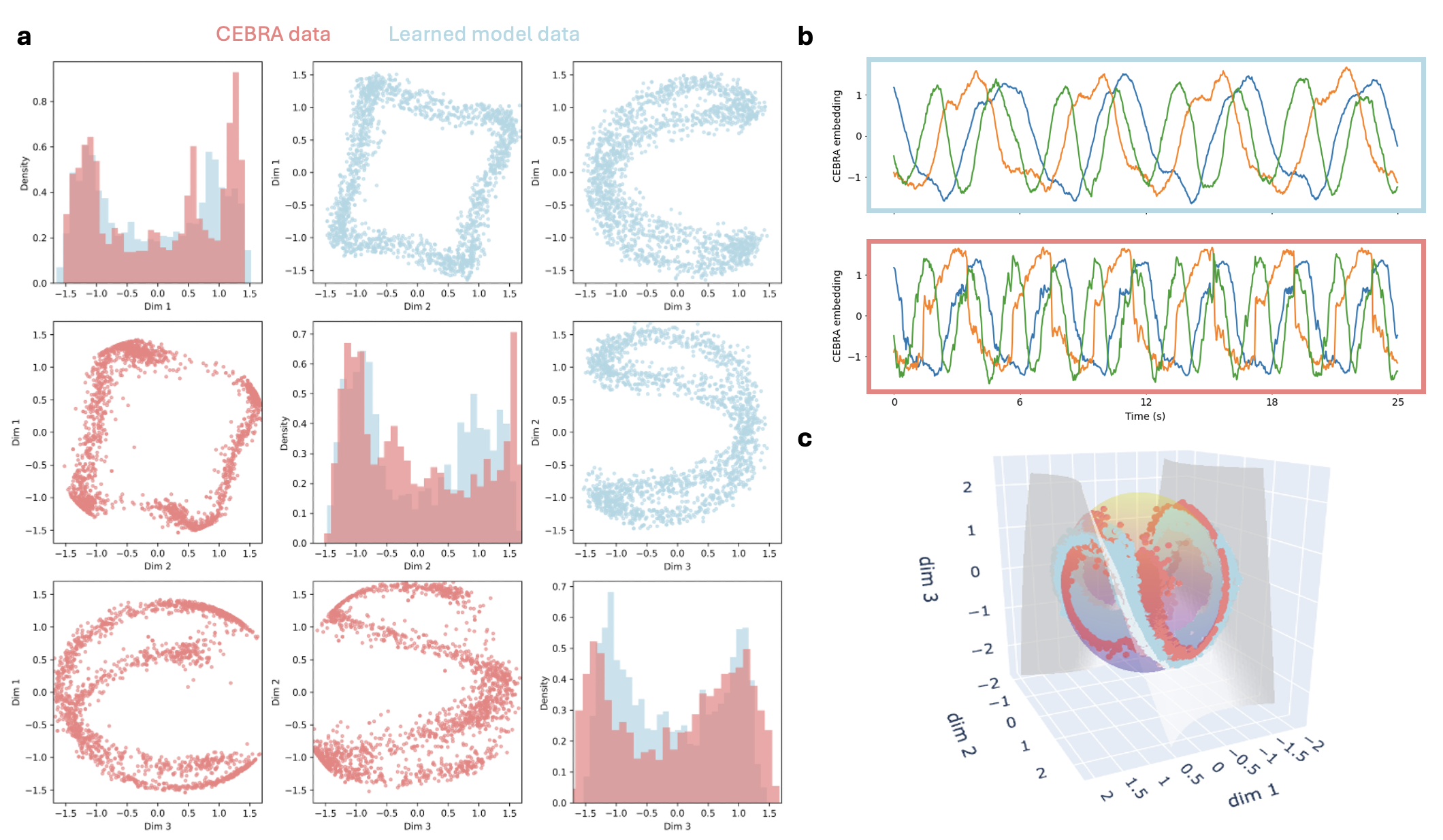}
\caption{\label{fig:mouse} 
Application of stochastic HN dynamics to a low-dimensional representation of neural data from a mouse hippocampus while moving along a track, trained with the position information of the mouse, using CEBRA~\cite{schneider2023learnable}. (a) Stationary distribution of the model resembles stationary distribution from the data. (b) Time series show similar behavior, although absolute frequencies differ due to behavioral changes from the mouse. (c) Intersecting Hamiltonian surfaces with points from the CEBRA data and model-generated data.
} 
\end{figure*}
We also apply our method to mouse hippocampus recordings while moving around a track~\cite{schneider2023learnable, grosmark2016diversity}.
We use low-dimensional embeddings from the CEBRA method trained with auxiliary behavioral variables of the mouse position and direction~\cite{schneider2023learnable}.
Our algorithm is similar to that used for the worm data in the main text, but now we do not use the data augmentation technique since we have more data points.
In addition, we explicitly account for the gradient dynamics in the data when fitting the curl component by subtracting out the gradient component before fitting.

\section{Scaling to larger dimensions}
Due to our use of Nambu Hamiltonians for the curl component, the method does not scale well to higher dimensions since Eq.~\eqref{eq:hamiltonian_dynamics} will become intractable.
We briefly describe another fitting method that is more scalable and enables the use of the full constraint from Eq.~\eqref{eq:general_constraint}.
First, the gradient component can be estimated from the stationary distribution.
The $\mathbf{g}$ component can be represented with a neural network, and the square of the value of the constraint in Eq.~\eqref{eq:general_constraint} is added to the loss with a regularization parameter. 
The first term of Eq.~\eqref{eq:general_constraint} is easy to calculate.
The second term can be hard to get, but we can get an unbiased estimate of it using the Skilling-Hutchinson estimator (similar to sliced score matching~\cite{song2020sliced}).
Then, the curl component can be optimized with the standard Euler-Maruyama maximum-likelihood loss function with the soft constraint on $\mathbf{g}$.

\clearpage

\bibliography{supp_Ref.bib}